\documentclass[prb,aps,epsf,twocolumn,longbibliography]{revtex4-1}
\usepackage{times}
\usepackage{graphicx}
\usepackage{float}
\usepackage{latexsym,amsmath,amssymb,bm,euscript}
\usepackage{xcolor}
\usepackage{epstopdf}
\usepackage[colorlinks=true,linkcolor=blue,citecolor=blue,urlcolor=blue]{hyperref}
\usepackage{hyperref}
\usepackage{amsmath}
\usepackage{ulem}

\begin{document}

\title{Chiral spin liquid and quantum phase diagram of spin-$\frac{1}{2}$ $J_1$-$J_2$-$J_{\chi}$ model on the square lattice}
\author{Xiao-Tian Zhang$^1$}
\author{Yixuan Huang$^{2,3}$}
\author{Han-Qing Wu$^4$}
\email{wuhanq3@mail.sysu.edu.cn}
\author{D. N. Sheng$^5$}
\email{donna.sheng1@csun.edu}
\author{Shou-Shu Gong$^6$}
\email{shoushu.gong@gbu.edu.cn}
\affiliation{
$^1$ School of Physics, Beihang University, Beijing 100191, China\\
$^2$ Theoretical Division, Los Alamos National Laboratory, Los Alamos, New Mexico 87545, USA \\
$^3$ Center for Integrated Nanotechnologies, Los Alamos National Laboratory, Los Alamos, New Mexico 87545, USA \\
$^4$ Guangdong Provincial Key Laboratory of Magnetoelectric Physics and Devices, School of Physics, Sun Yat-sen University, Guangzhou, 510275, China \\
$^5$ Department of Physics and Astronomy, California State University Northridge, Northridge, California 91330, USA\\
$^6$ School of Physical Sciences, Great Bay University, Dongguan 523000, China, and \\
Great Bay Institute for Advanced Study, Dongguan 523000, China
}

\begin{abstract}

We study the spin-$1/2$ Heisenberg model on the square lattice with the first- and second-nearest-neighbor antiferromagnetic couplings $J_1$ and $J_2$, as well as the three-spin scalar chiral coupling $J_{\chi}$. Using density matrix renormalization group calculations, we obtain a quantum phase diagram of this system for $0 \leq J_2/J_1 \leq 1.0$ and $0 \leq J_{\chi}/J_1 \leq 1.5$.
We identify the N\'eel and stripe magnetic order phase at small-$J_{\chi}$ coupling.
With growing $J_{\chi}$, we identify the emergent chiral spin liquid (CSL) phase characterized by the quantized spin Chern number $C = 1/2$ and entanglement spectrum with the quasidegenerate group of levels agreeing with chiral SU(2)$_1$ conformal field theory, which is an analog of the $\nu = 1/2$ Laughlin state in the spin system.
In the vicinity of the N\'eel and CSL phase boundary, our numerical results do not find evidence to support the coexistence of N\'eel order and topological order that was conjectured by mean-field calculations.
In the larger-$J_2$ and -$J_{\chi}$ coupling regime, the entanglement spectrum of the ground state also exhibits the chiral quasidegeneracy consistent with a CSL, but the adiabatic flux insertion simulations fail to obtain the quantized Chern number.
By analyzing the finite-size scaling of the magnetic order parameter, we find the vanished magnetic order suggesting a magnetic disorder phase, whose nature needs further studies.
Different from the spin-$1$ $J_1$-$J_2$-$J_\chi$ model, we do not find the coexistent stripe magnetic order and topological order.
We also investigate the $J_{\chi}$ dominant regime and find a strong tendency of the system to develop a dimer order rather than the chiral spin magnetic order observed in the spin-$1$ model.
Our results unveil interesting quantum phases in this spin-$1/2$ model and also demonstrate the drastic differences between the spin-$1/2$ and spin-$1$ cases.
\end{abstract}

\maketitle

\section{Introduction}
 
Quantum spin liquids (QSLs) are exotic states of matter in frustrated quantum magnets~\cite{Savary_2016,Zhou_2017,Broholm_2020,lancaster2023quantum,Kivelson_2023}, which can escape from forming conventional orders even at zero temperature.  Remarkably, QSLs have long-range quantum entanglement as well as fractionalized excitations~\cite{wen1990, Wen1991,Senthil2000,Senthil2001,chen2010}.
Among the various types of QSLs, the chiral spin liquid (CSL) is a specific class of QSL that is an analog of the fractional quantum Hall state~\cite{Kalmeyer1987, wen1989,Baskaran1989,Yang1993,Haldan1995}.
CSLs break time-reversal and parity symmetries but preserve lattice translational and spin rotational symmetries.
Kalmeyer and Laughlin first proposed that, in a time-reversal-invariant spin model with geometric frustration, one may realize a CSL as the analog of the $\nu = 1/2$ Laughlin state through spontaneous time-reversal-symmetry breaking~\cite{Kalmeyer1987}.
Later, based on the wave functions of fractional quantum Hall states, the parent Hamiltonians for which the CSLs are the exact ground states have also been constructed in spin models with long-range interactions~\cite{Ronny2007,Ronny2009,Greiter2014}.

More interestingly, doping a CSL may lead to exotic anyon superconductivity~\cite{wen1989,wilczek1990fractional}.
After extensive search for decades, this Kalmeyer-Laughlin CSL has been identified not only in various spin-$1/2$ systems with either geometric frustration or competing interactions~\cite{He2014,Gong_2014,Wietek2015,Hu2015,Gong2015,Hickey2017,Gong2017,Wietek2017,nielsen2013,Didier2017,Chen2021,Hasik2022}, but also in the Hubbard model near the Mott transition~\cite{Szasz2020,Boos2020,Szasz2021,Cookmeyer2021,Chen2022}. 
Very recently, unbiased density matrix renormalization group (DMRG) studies have also established a ($d+id$)-wave topological superconducting phase near the doped CSL regime~\cite{Jiang2020,Huang2022,Huang2023}.

On the other hand, CSL may also play an important role in cuprate superconductors. Recently, thermal Hall measurements for cuprate superconductors have found a giant thermal Hall conductivity at small doping level even including the half-filled case~\cite{Grissonnanche_2019,grissonnanche2020chiral,boulanger2020thermal,boulanger2022thermal,boulanger2023thermal}, which triggered the theoretical interests in the origin of thermal Hall conductivity~\cite{samajdar2019enhanced,samajdar2019thermal,Han2019,li2019thermal,li2019theory,Zhang2021,merino2022majorana,Varma2020,Guo2021,ye2021phonon,Mangeolle2022PRX,Mangeolle2022,Flebus2022,Sun2022}.
Based on mean-field calculations, it was proposed that there is a coexistent regime of N\'eel magnetic order and CSL topological order~\cite{samajdar2019enhanced,Zhang2021} in the spin-$1/2$ square-lattice Heisenberg model with the nearest-neighbor (NN) $J_1$ and next-nearest-neighbor (NNN) $J_2$ antiferromagnetic (AFM) exchange interactions, as well as the three-spin scalar chiral coupling $J_\chi$ describing the effect of external magnetic fields in the thermal Hall experiments.  
The coexistent N\'eel AFM order and neutral spinon excitations may provide a natural understanding of the experimental observations~\cite{samajdar2019enhanced,Zhang2021}.

In previous studies on this microscopic spin model, the N\'eel AFM order and Kalmeyer-Laughlin CSL state have been found~\cite{merino2022majorana,nielsen2013,Liu_2016},
but the possible coexistence has not been examined in numerical calculations.
Very recently, the similar spin-$1$ $J_1$-$J_2$-$J_{\chi}$ model has been comprehensively studied by DMRG calculations, which find no evidence for coexistent N\'eel order and Abelian topological order but unveil a coexistence of stripe AFM order and a non-Abelian topological order in the larger-$J_2$ regime~\cite{huang2021}.
This coexistence is characterized by a strong magnetic order and the entanglement spectrum that agrees with the expected CSL state~\cite{huang2021}.
These reported results of the spin-$1$ model also stimulate an interest in the similar phenomenon in the corresponding spin-$1/2$ system.

In this work, we examine the quantum phase diagram of the spin-$1/2$ $J_1$-$J_2$-$J_\chi$ model on the square lattice (see the schematic figure on the left top corner of Fig.~\ref{fig:model}) by means of DMRG calculations, with the focus on exploring the possible coexistence of CSL topological order and magnetic order.
In the studied parameter regime $0 \leq J_2/J_1 \leq 1.0$ and $0 \leq J_{\chi} / J_1 \leq 1.5$, we identify a N\'eel AFM phase, a stripe AFM phase, a CSL phase, and a magnetic disorder regime, as shown in Fig.~\ref{fig:model}. 
We confirm the CSL state by showing the quantized topological entanglement entropy $\gamma \simeq (\ln 2)/2$, chiral entanglement spectrum counting $\{ 1, 1, 2, 3, 5,...\}$, and quantized spin Chern number $C = 1/2$.
Furthermore, we carefully investigate the phase transition from the N\'eel AFM phase to the CSL with growing chiral coupling.
While the transition can be identified by various quantities, we do not find evidence to support a coexistence of topological order and N\'eel order near the phase boundary. 

On the other hand, at the larger-$J_2$ side we find either the CSL state or a magnetic disorder state with increased $J_{\chi}$ coupling.
While the entanglement spectrum of the ground state in the magnetic disorder regime can exhibit the quasidegenerate group of levels agreeing with chiral SU(2)$_1$ conformal field theory, the adiabatic flux insertion simulations fail to obtain a half-quantized spin Chern number in the studied system size. 
Meanwhile, the finite-size extrapolation of magnetic order parameters strongly suggests the absent magnetic order in this regime.
Therefore, our DMRG results indicate that the coexistence of topological order and stripe AFM order observed in the spin-$1$ model~\cite{huang2021} may not exist in the studied spin-$1/2$ case. 
In addition, we examine the $J_{\chi}$-dominant regime where a chiral spin state (CSS) magnetic order~\cite{rabson1995} (see Appendix~\ref{Appendix A:Classical phase diagram}) has been found in the spin-$1$ $J_1$-$J_2$-$J_{\chi}$ model~\cite{huang2021}.
Nonetheless, we do not find the CSS magnetic order but a strong tendency of the system to develop a dimer order spontaneously breaking lattice translational symmetry.

Compared with the spin-$1$ model, our results unveil the big differences between the spin-$1/2$ and spin-$1$ $J_1$-$J_2$-$J_{\chi}$ models, showing the important role of quantum fluctuations in the emergence of the different quantum states. 
In particular, the existence and absence of phase coexistence in the spin-$1$ and spin-$1/2$ systems may shed light on further investigation of the coexisting topological order and conventional order.
We also discuss the potential nature of the magnetic disorder regime, which we leave for future study.

The organization of our paper is as follows:
In Sec.~\ref{sec:method}, we introduce the model and the details of  DMRG calculations. 
In Sec.~\ref{sec:Phase diagram 1}, we identify the phase diagram at the smaller $J_2$ side. 
We characterize the CSL state and show the phase transition from the N\'eel to the CSL phase. 
We also explore the possible phase coexistence between the two phases.
In Sec.~\ref{sec:Phase diagram 2}, we discuss the quantum phases at the larger-$J_2$ side and carefully examine the properties of the magnetic disorder regime.
In Sec.~\ref{sec:VBC}, we go beyond the phase diagram in Fig.~\ref{fig:model} and discuss the possible dimer order phase in the $J_\chi$-dominant regime.
We summarize the results is Sec.~\ref{sec:Summary and discussion}.

\section{Model Hamiltonian and methods}
\label{sec:method}

\begin{figure}[ht]
\includegraphics[width = 1.0\linewidth]{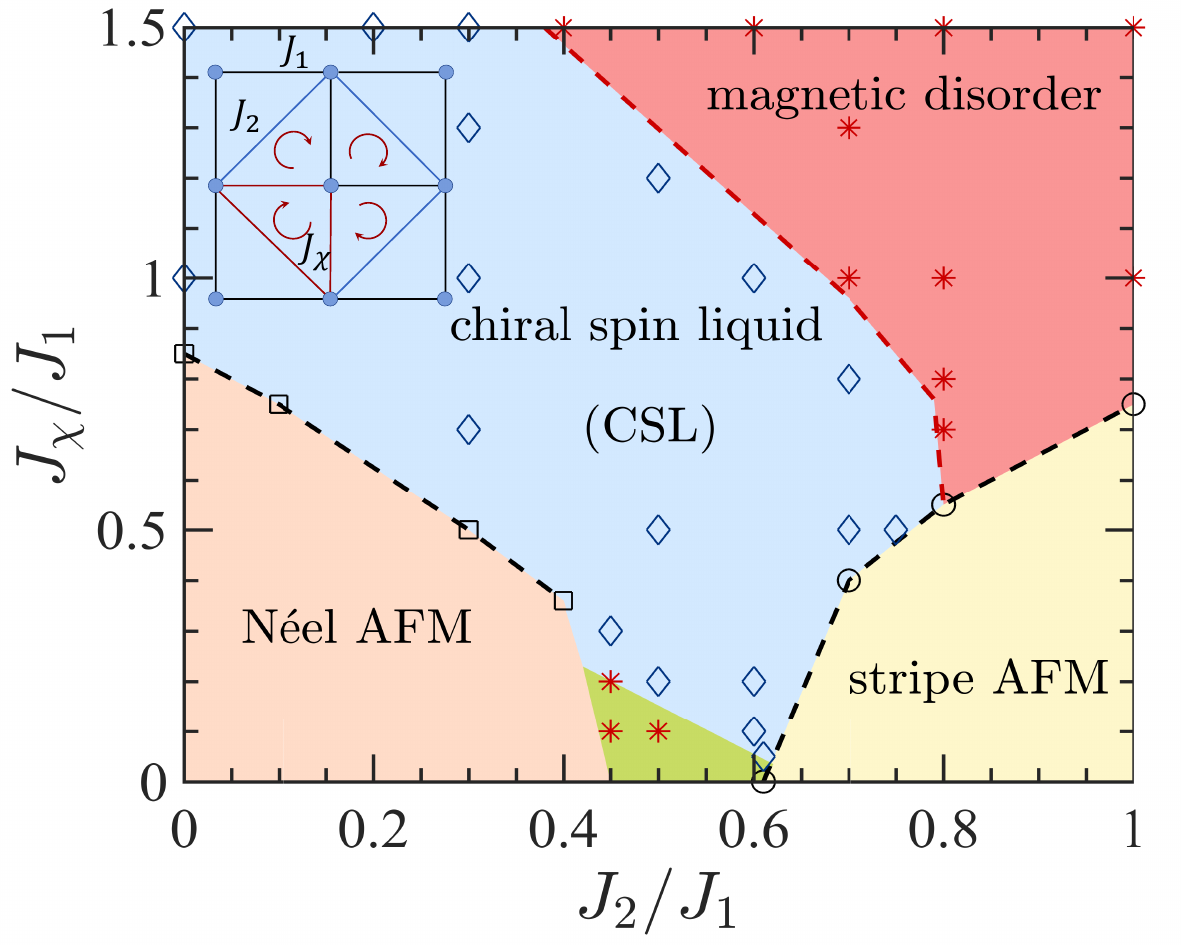}
\caption{Model illustration and phase diagram.
The inset on the left top corner is a schematic illustration of the spin-$1/2$ $J_1$-$J_2$-$J_\chi$ Heisenberg model on the square lattice. $J_1$ and $J_2$ are the NN and NNN AFM interactions, respectively. $J_{\chi}$ is the magnitude of the three-spin chiral couplings ${\bf S}_i \cdot ({\bf S}_j \times {\bf S}_k)$ in the Hamiltonian in Eq.~\eqref{eq:ham}, which includes all the four kinds of triangles in each plaquette. For each triangle, the sites $i, j$ and $k$ follow the same clockwise direction.
The quantum phase diagram of the spin-$1/2$ $J_1$-$J_2$-$J_\chi$ model is obtained with $0 \leq J_2/J_1 \leq 1.0$ and $0 \leq J_\chi/J_1 \leq 1.5$. With tuning couplings, we identify the N\'eel AFM, stripe AFM, chiral spin liquid, and a magnetic disorder regime. The blue diamonds indicate the parameter points in which we can obtain a quantized spin Chern number $C = 1/2$ by the adiabatic flux insertion simulations on the $L_y = 8$ systems. For the intermediate $J_2/J_1$ regime with small-$J_{\chi}$ couplings, the red asterisks represent the parameter points where our DMRG simulations do not obtain a quantized Chern number. In the magnetic disorder regime, the entanglement spectrum of the ground state shows the chiral counting consistent with a chiral spin liquid, but the flux insertion simulations fail to obtain a quantized Chern number.}
\label{fig:model}
\end{figure}

We consider the spin-$1/2$ square-lattice model with the NN ($J_1$) and NNN ($J_2$) AFM Heisenberg interactions, as well as the three-spin scalar chiral coupling $J_{\chi}$ for all four small triangles in each plaquette of the square lattice.
The Hamiltonian is defined as
\begin{equation}
H = \sum_{\langle i, j \rangle} J_1 {\bf S}_i \cdot {\bf S}_j + \sum_{\langle\langle i, j \rangle\rangle} J_2 {\bf S}_i \cdot {\bf S}_j+\sum_{\bigtriangleup} J_{\chi} {\bf S}_i \cdot ({\bf S}_j \times {\bf S}_k),
\label{eq:ham}
\end{equation}
where the sums ${\langle i, j \rangle}$ and $ {\langle\langle i, j \rangle\rangle}$ run over all the NN and NNN bonds, respectively. 
The chiral couplings involve the sum over all four kinds of triangles in each plaquette, where the sites $i, j$ and $k$ follow the same clockwise direction as shown in Fig.~\ref{fig:model}.

\begin{figure*}[ht]
\includegraphics[width = 0.32\linewidth]{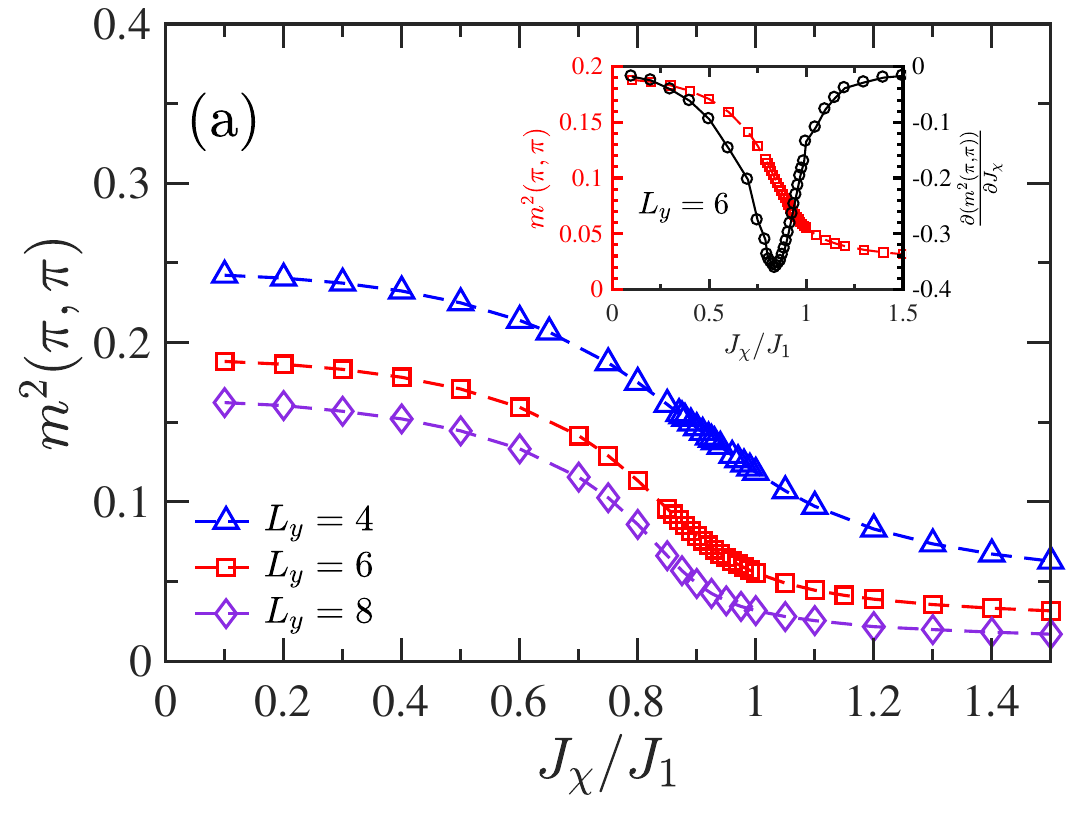}
\includegraphics[width = 0.33\linewidth]{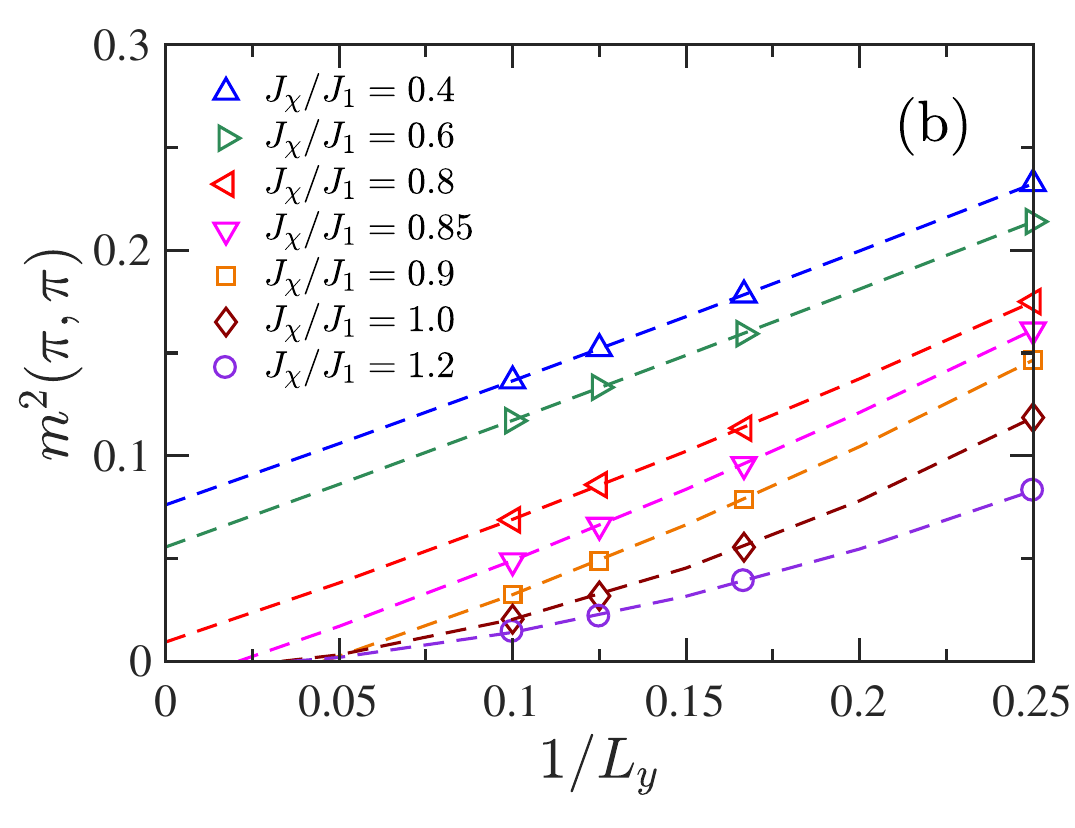}
\includegraphics[width = 0.32\linewidth]{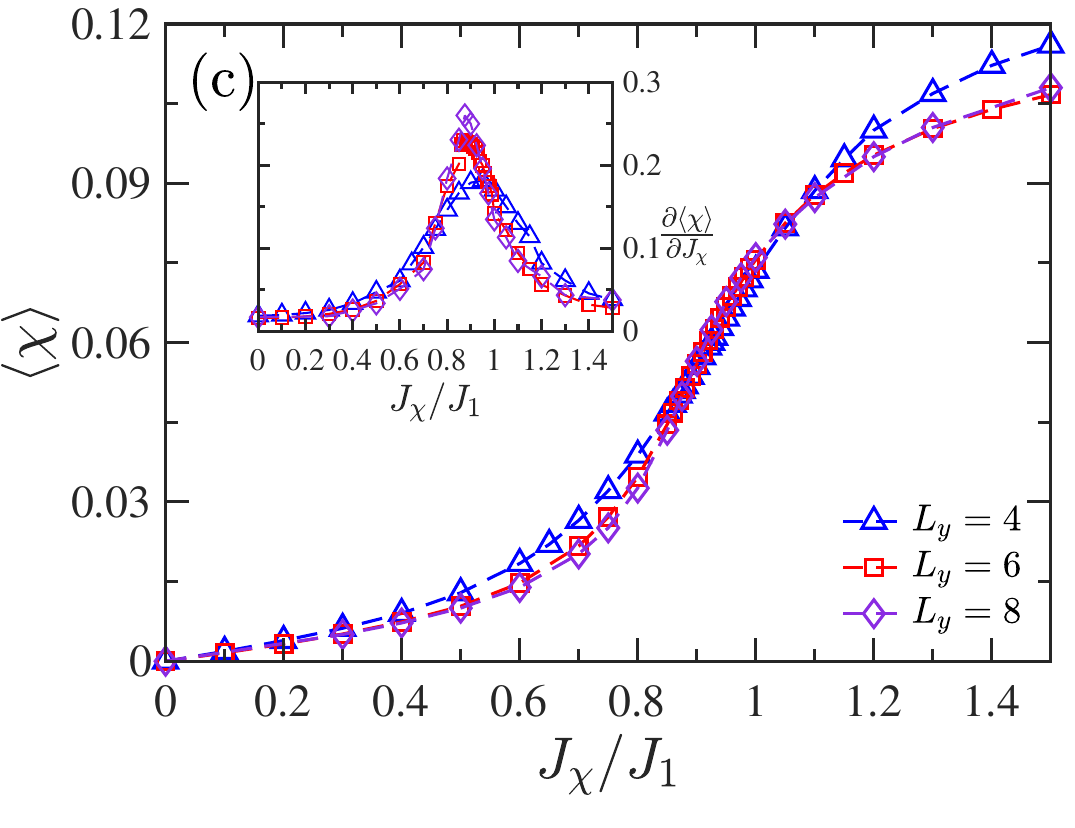}
\caption{Quantum phase transition with vanishing the N\'eel AFM order by chiral coupling at $J_2 = 0$.
(a) N\'eel AFM order parameter $m^2(\pi,\pi)$ versus $J_{\chi} / J_1$. The data of $m^2(\pi,\pi )$ are obtained from the middle $ L_y \times L_y$  sites of a long cylinder system with $L_y=4,6,8$. The inset is the $J_{\chi} / J_1$ dependence of $m^2(\pi,\pi )$ and its first-order derivative for $L_y = 6$. (b)  Finite-size scaling of $m^2(\pi,\pi )$ with $L_y = 4, 6, 8, 10$ for different $J_{\chi} / J_1$. The dashed lines denote the polynomial fittings up to the second order of $1/L_y$. (c) Scalar chiral order $\langle \chi \rangle$ versus $J_{\chi}/J_1$ on different system sizes. The chiral order defined as $\langle \chi \rangle \equiv \langle {\bf S}_i \cdot ({\bf S}_j \times {\bf S}_k) \rangle$ is measured in the bulk of the long cylinder and is uniform. The four kinds of triangles in each plaquette have the same magnitude of chiral order. The inset shows the first-order derivative of the chiral order with respect to $J_\chi/J_1$, which is equivalent to the second-order derivative of ground-state energy. The symbols in figure (c) and its inset follow the same definitions.}
\label{fig:CSL}
\end{figure*}

We determine the quantum phases of the model by using DMRG~\cite{White1992} calculations.
In DMRG simulation, we study the system on the cylinder geometry with the periodic boundary conditions along the circumference direction ($y$ direction) and the open boundary conditions along the axis direction ($x$ direction).
We denote $L_y$ and $L_x$ as the site numbers in the two directions, respectively.
We implement the spin SU(2) symmetry~\cite{McCulloch2002} in the finite DMRG simulations, which allows us to study the systems with circumference up to $L_y = 12$.
We keep the bond dimensions up to $6000$ SU(2) multiplets, which are equivalent to about $24000$ U(1) states.
For the $L_y = 12$ systems, the DMRG truncation error is smaller than $5 \times 10^{-5}$, while for smaller $L_y$ systems, we can obtain very accurate results with the truncation error smaller than $1 \times 10^{-6}$.
In addition, we use the TeNPy~\cite{tenpy} package to perform the infinite DMRG simulations on systems with the circumference up to $L_y = 8$, in which we keep the bond dimensions up to $3000$ U(1) states, with the truncation error near $10^{-5}$.

We obtain the quantum phase diagram of the model at $0 \leq J_2/J_1 \leq 1.0$ and $0 \leq J_{\chi}/J_1 \leq 1.5$, as shown in Fig.~\ref{fig:model}. 
We set $J_1 = 1.0$ as the energy unit.
In the presence of a moderate chiral coupling, we identify two magnetic order phases, a CSL phase, and a magnetic disorder regime in the studied parameter region.
For the intermediate nonmagnetic regime $J_2/J_1 \sim 0.55$, we find that the CSL state can be easily driven by a small chiral coupling. 

\section{N\'eel order, chiral spin liquid, and the phase transition}
\label{sec:Phase diagram 1}

In this section, we will demonstrate our numerical results to identify the N\'eel AFM phase, CSL phase, and the phase transition.
We will also discuss our investigation of the possible phase coexistence.

\subsection{N\'eel phase and the phase transition}

For $J_2 = J_{\chi} = 0$, the system possesses a N\'eel AFM order, which would be suppressed by the increased $J_{\chi}$ coupling.
With growing chiral coupling, a transition from the N\'eel order to the CSL has been proposed at $J_\chi / J_1 \simeq 0.189$ by exact diagonalization calculation~\cite{nielsen2013}, showing that the AFM order will be easily melted by chiral coupling. 
On the other hand, a recent exact diagonalization calculation of quantum fidelity on a $4 \times 4$ cluster suggests the transition at $J_\chi / J_1 \simeq 0.7$~\cite{merino2022majorana}, and a spinon mean-field study finds the transition at $J_\chi / J_1 \simeq 1.35$~\cite{Liu_2016}.

To accurately determine the phase transition with vanishing N\'eel order, we first compute the N\'eel order parameter $m^2(\pi,\pi)$ defined as 
\begin{equation}
m^2(\vec{k}) = \frac{1}{N^{2}_m}\sum_{i,j}\langle {\bf S}_i \cdot {\bf S}_j \rangle e^{i\vec{k}\cdot(\vec r_i-\vec r_j)}
\end{equation}
at the momentum $\vec{k} = (\pi, \pi)$.
In the summation of the above equation, we only consider the middle $N_m = L_y \times L_y$ sites on a long cylinder to avoid the edge effects. 
In Fig.~\ref{fig:CSL}(a), we show $m^2(\pi,\pi)$ with growing $J_\chi/J_1$ for $L_y = 4, 6, 8$ ($J_2 = 0$), which all have a behavior change at $J_\chi/J_1 \simeq 0.85$ that can be clearly identified in the derivative as shown in the inset of Fig.~\ref{fig:CSL}(a).
In Fig.~\ref{fig:CSL}(b), we further analyze the finite-size scaling of $m^2(\pi,\pi)$ with $L_y = 4,6,8,10$, which are extrapolated using polynomial fitting up to the second order of $1/L_y$.
The extrapolated results support a transition with vanished N\'eel AFM order at $J_\chi/J_1 \simeq 0.85$, which is consistent with the finding in Fig.~\ref{fig:CSL}(a).

Since the phase transition may be characterized by the ground-state energy, we also study the scalar chiral order parameter $\langle \chi \rangle$ in the bulk of the cylinder versus $J_\chi/J_1$.
The chiral order is defined as $\langle \chi \rangle \equiv \langle {\bf S}_i \cdot ({\bf S}_j \times {\bf S}_k) \rangle$, where the sites $i,j$ and $k$ follow the definition of the small triangle shown in the inset of Fig.~\ref{fig:model}.
Thus, this chiral order is equivalent to the first-order derivative of the ground-state energy with respect to the chiral coupling.
The obtained chiral orders on different system sizes [Fig.~\ref{fig:CSL}(c)] and their derivatives shown in the inset also characterize a transition at $J_\chi/J_1 \simeq 0.85$.
Therefore, our DMRG results indicate that a chiral coupling as large as the Heisenberg interaction $J_1$ is required to suppress the N\'eel order.
In the same way, we can determine the phase boundary with vanishing the N\'eel order in Fig.~\ref{fig:model}. 
More data can be found in Fig.~\ref{appfig:J_nn_01,J_nn_03,J_nn_04} 
 (see Appendix~\ref{Appendix B:Quantum phase transition}).

\begin{figure*}[ht]
\includegraphics[width = 0.33\linewidth]{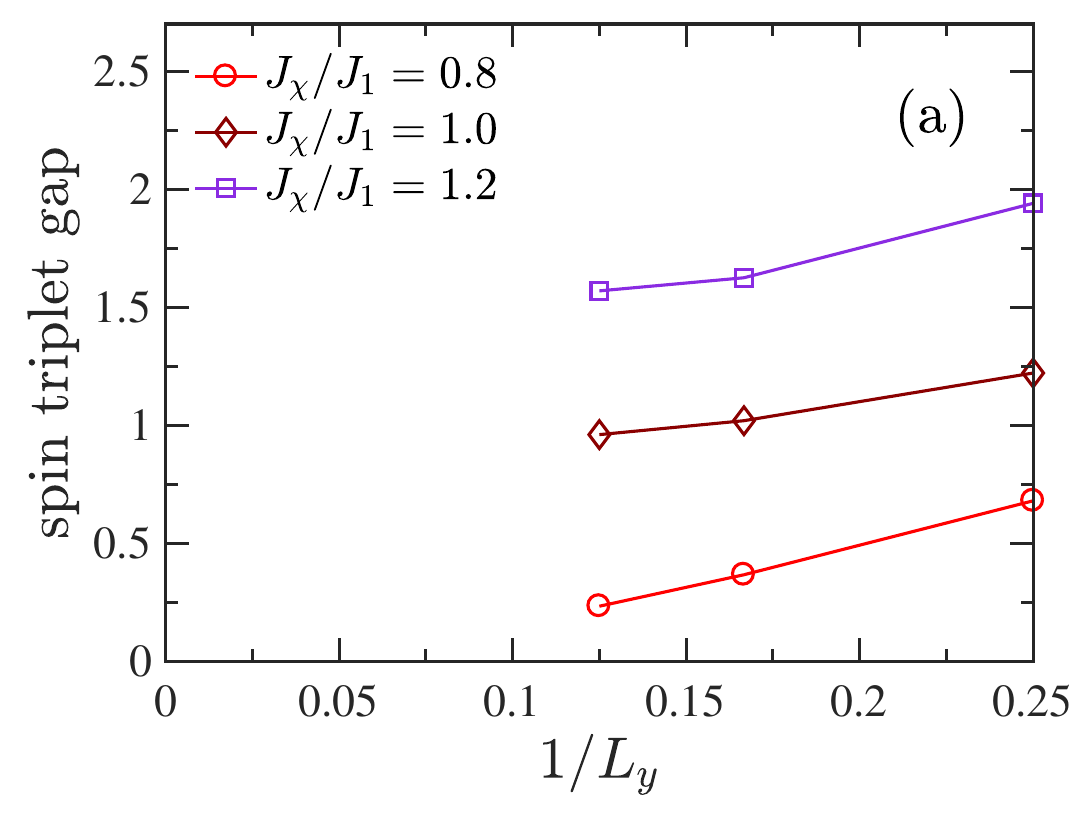}
\includegraphics[width = 0.33\linewidth]{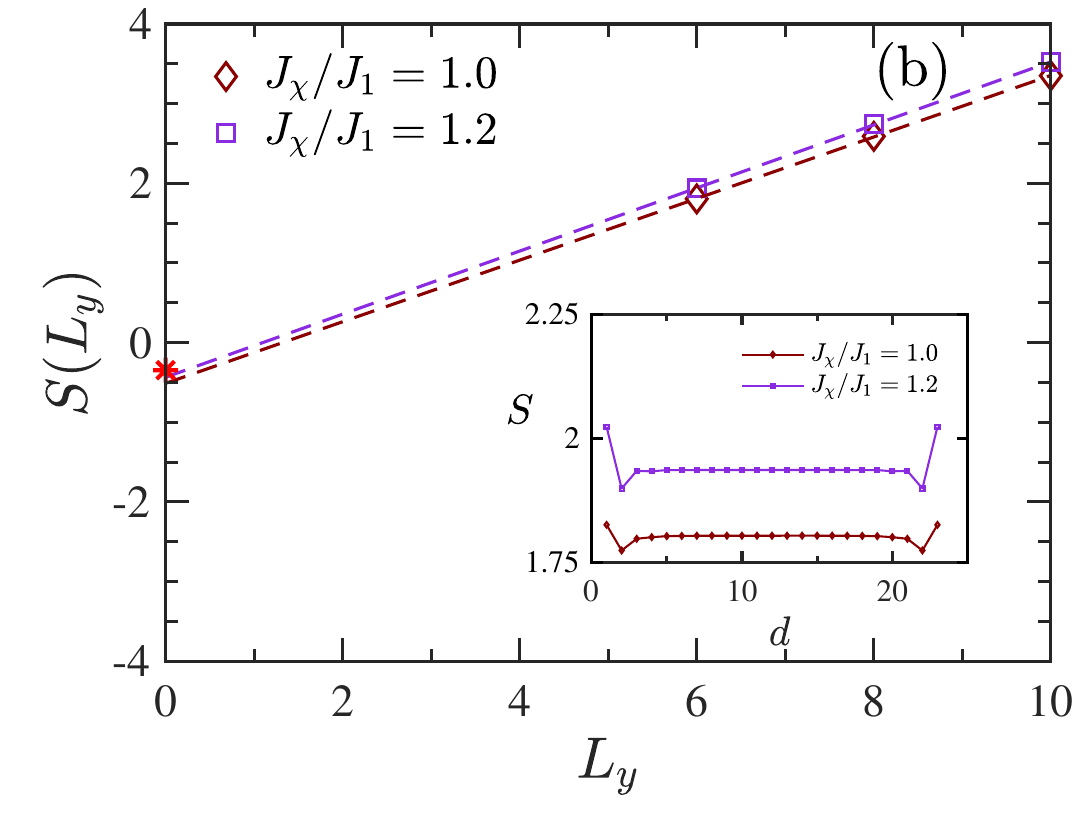}
\includegraphics[width = 0.33\linewidth]{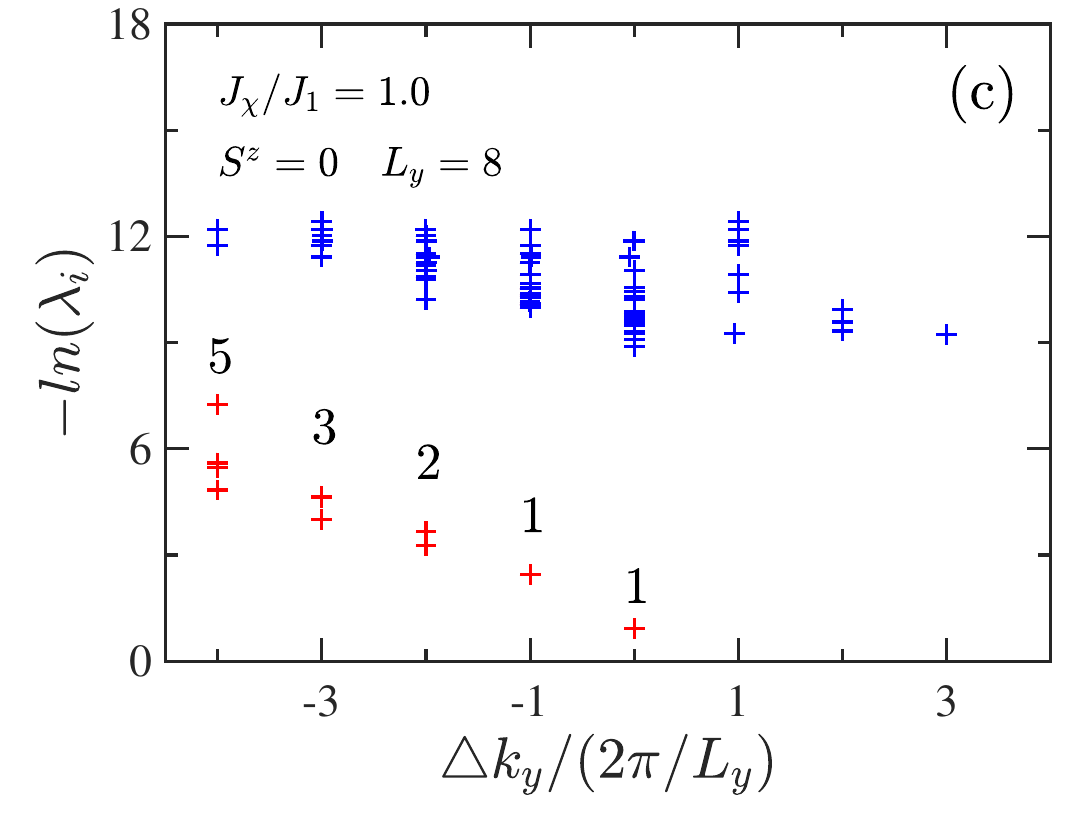}
\caption{Identification of the chiral spin liquid state at $J_2 = 0$.
(a) Finite-size scaling of the spin-triplet gap, which is obtained for the middle $L_y \times L_y$ sites on the $L_y=4,6,8$ systems. We show the results in both N\'eel AFM phase ($J_{\chi} / J_1 = 0.8$) and chiral spin liquid phase ($J_{\chi} / J_1 = 1.0, 1.2$), which are clearly extrapolated to vanished and finite, respectively.
(b) Entanglement entropy $S$ versus cylinder circumference $L_y$ at $J_\chi / J_1 = 1.0, 1.2$. We fit the data by the formula $S = a L_y - \gamma$. The red asterisk shows the theoretical prediction $\gamma =  (\ln 2) / 2$ for the $\nu = 1/2$ Laughlin state. The inset shows the entanglement entropy obtained for different subsystem length $d$, which follows the area law of entropy.
(c) Entanglement spectrum labeled by the quantum numbers of total spin $S^z = 0$ and relative momentum along the $y$ direction $\triangle{k_y}$. $\lambda_i$ is the eigenvalue of reduced density matrix. We show the results for $J_\chi / J_1 = 1.0$ at the $L_y=8$ system. The numbers $\{ 1, 1, 2, 3, 5\}$ denote the near-degenerate low-lying levels separated by an entanglement gap from higher levels, which agrees with the prediction of the SU(2)$_1$ conformal field theory. In the figure, the smaller eigenvalues are not shown.}
\label{fig:gap_TEE_entanglement}
\end{figure*}

\subsection{Identification of the chiral spin liquid}

With suppressed N\'eel order by increasing the chiral coupling, we can identify an emergent gapped Kalmeyer-Laughlin CSL by DMRG calculations, which is equivalent to the $\nu = 1/2$ bosonic fractional quantum Hall state~\cite{Kalmeyer1987}.
First of all, we calculate the spin-triplet gap.
We obtain the gap in the bulk by calculating the lowest-energy states in the total spin $S = 0$ and $S = 1$ sectors for the middle $L_y \times L_y$ sites by DMRG.
We first calculate the ground state in the $S = 0$ sector on a long cylinder and then sweep the $S = 1$ sector for the middle $L_y \times L_y$ sites, which can avoid the edge excitations~\cite{yan2011spin}.
In Fig.~\ref{fig:gap_TEE_entanglement}(a), we show the system size dependence of the triplet gap obtained on the $L_y = 4,6,8$ systems at different $J_{\chi}$ and $J_2 = 0$.
One can see that the gaps drop fast versus $1/L_y$ and go to zero in the thermodynamic limit at $ J_\chi / J_1 = 0.8$, consistent with the N\'eel AFM order with gapless Goldstone modes.
In contrast, for $J_\chi/J_1 = 1.0$ and $1.2$, the gaps are extrapolated to finite values, showing the gapped spin-triplet excitations.

Next, we identify the quantized topological entanglement entropy of the Kalmeyer-Laughlin CSL by analyzing the circumference dependence of entanglement entropy~\cite{Jiang2012}.
For gapped topological states, the entanglement entropy follows the area law and thus mainly depends on the system circumference $L_y$~\cite{Eisert2010}.
In particular, the entropy $S(L_y)$ of the minimum entangled state with a smooth boundary of length $L_y$ will take the form
\begin{equation}
S(L_y) \simeq a L_y - \gamma,
\label{eq:TEE}
\end{equation}
where $\gamma$ is the topological entanglement entropy, which is a universal additive constant characterizing the long-range entanglement in the ground state and is related to the total quantum dimension $D$ of a topological order as $\gamma = \ln D$~\cite{Kitaev2006,Wen2006}.
The coefficient $a$ is nonuniversal due to the short-distance physics near the boundary.
In DMRG calculation, it is straightforward to obtain the entanglement entropy by bipartitioning the whole system into two subsystems.
The entropy can be obtained as $S = -\sum_{i} \lambda_i \ln \lambda_i$, where $\lambda_i$ are the eigenvalues of the reduced density matrix of the subsystems.
In the inset of Fig.~\ref{fig:gap_TEE_entanglement}(b), the entropy versus subsystem length shows that the entropy is almost independent of the subsystem length, which agrees with the area law of entanglement entropy for such a quasi-one-dimensional gapped system~\cite{Eisert2010}.
Since DMRG simulation on topological order states would naturally find the minimum entangled state~\cite{Jiang2012}, in Fig.~\ref{fig:gap_TEE_entanglement}(b), we can study the scaling of the obtained entropy $S(L_y)$ versus $L_y$ ($L_y = 6, 8, 10$) for $J_\chi / J_1 = 1.0$ and $1.2$.  
Notice that the entropy data for each $L_y$ are obtained on long cylinders, which is almost independent of cylinder length and can be taken as the result in the $L_x \rightarrow \infty$ limit.
Following Eq.~\eqref{eq:TEE}, we fit our entropy data linearly with the red asterisk showing the theoretical prediction $\gamma = (\ln 2) / 2$ for the Kalmeyer-Laughlin CSL~\cite{Wen2006}. 
One can see that although the finite-size effects may exist, the extrapolated $\gamma$ at $J_\chi / J_1 = 1.0, 1.2$ are very close to $(\ln 2) / 2$, which highly agrees with the Kalmeyer-Laughlin CSL state. 

Furthermore, we calculate the entanglement spectrum to characterize the CSL state because the entanglement spectrum has a one-to-one correspondence with the physical edge spectrum~\cite{Haldane2008,Qi2012}.
We use the quantum number total $S^z$ of the half system and the relative momentum quantum number along the $y$ direction, $\triangle{k_y}$ to label the obtained eigenvalues of the reduced density matrix~\cite{Cincio2013,Zaletel2013}. 
In Fig.~\ref{fig:gap_TEE_entanglement}(c), we present the entanglement spectrum of the ground state at $J_\chi/J_1 = 1.0$ on the $L_y = 8$ cylinder, which is obtained for the total $S^z = 0$ sector.
The spectrum clearly shows the quasidegenerate pattern $\{ 1, 1, 2, 3, 5,...\}$ with increasing $\triangle{k_y}$, and the quasidegenerate levels are well separated from the higher levels by an entanglement gap.
Such a chiral structure of entanglement spectrum agrees with the SU(2)$_1$ conformal field theory prediction and supports a Kalmeyer-Laughlin CSL~\cite{DiFrancesco:1997nk}.

\begin{figure}[h]
\includegraphics[width = 0.495\linewidth]{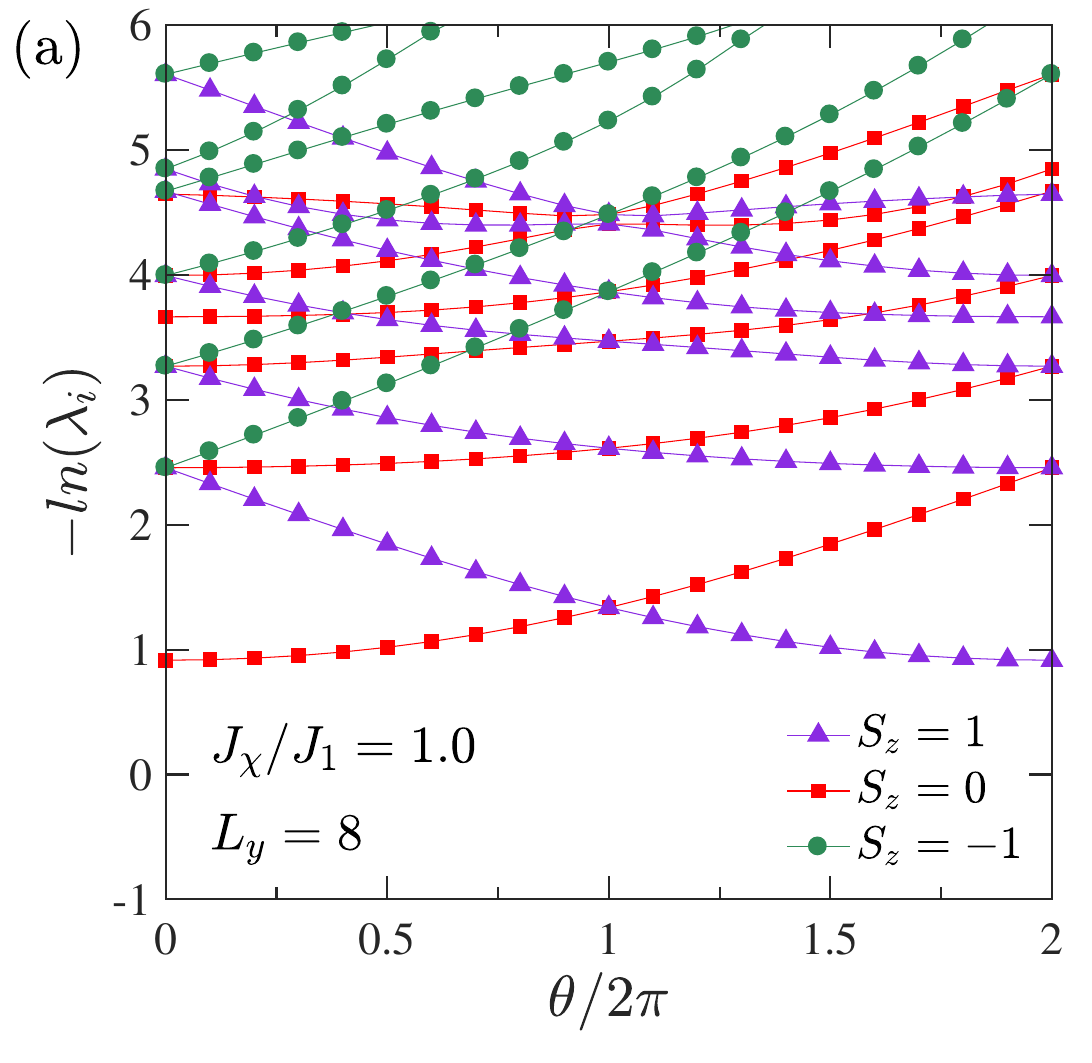}
\includegraphics[width = 0.495\linewidth]{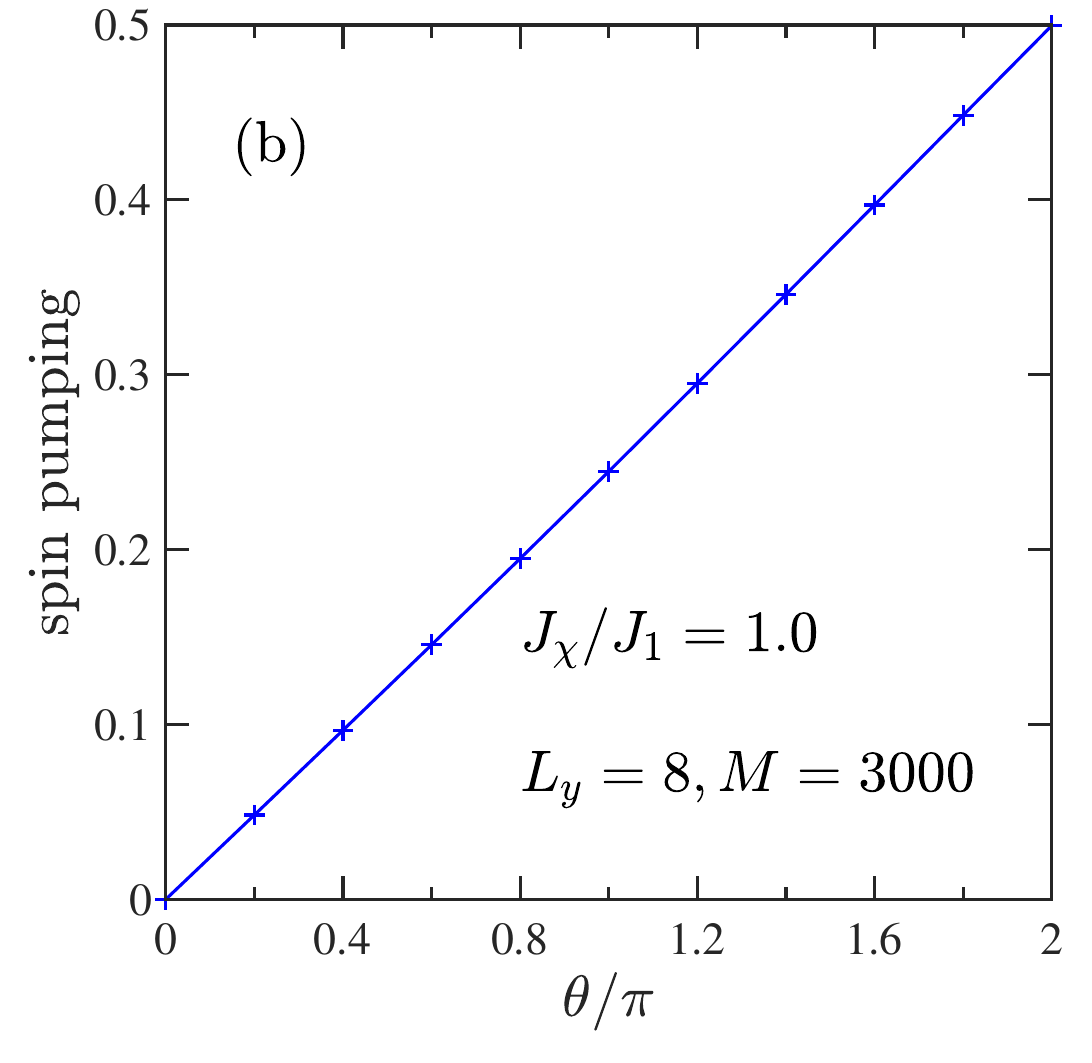}
\caption{Adiabatic flux insertion simulation and quantized spin Chern number for $J_\chi / J_1 = 1.0$, $J_2 = 0$ on the $L_y = 8$ cylinder.
(a) Entanglement spectrum flow with adiabatically inserted flux $\theta$, which is obtained by iDMRG simulation with $1000$ bond dimensions. The eigenvalues in the $S_z = 0, \pm 1$ sectors are labeled by different symbols. 
(b) Spin pumping with adiabatically inserted flux $\theta$, which is obtained by iDMRG simulation with $3000$ bond dimensions. In a period of inserted flux $\theta = 0 \rightarrow 2\pi$, a quantized magnetization moment $0.5$ is accumulated at one open boundary of the cylinder, indicating a fractionally quantized spin Chern number $C = 1/2$.}
\label{fig:ES_flow}
\end{figure}

At last, we compute the spin Chern number through adiabatic flux insertion and the change of entanglement spectrum with increased flux~\cite{Haldan1995,Gong_2014}.
Since inserting a flux along the cylinder is equivalent to imposing the twist boundary conditions in the circumference ($y$) direction, we consider the flux by replacing the spin flipping term $S^{+}_i S^{-}_j + H.c.$ to $S^{+}_i S^{-}_j e^{i \theta} + H.c.$ for these Heisenberg interacting bonds crossing the $y$ boundary.
For the chiral interactions crossing the $y$ boundary, we introduce the similar changes in the Hamiltonian terms.
We take one component $S^{+}_i S^{-}_j S^{z}_k + H.c.$ as an example, which is replaced as $S^{+}_i S^{-}_j S^{z}_k e^{i \theta} + H.c.$ in the presence of flux.
We adiabatically increase the flux by taking the obtained ground state at a given flux $\theta$ as the initial state for the next-step simulation of the added flux $\theta + \Delta \theta$.
The entanglement spectrum versus the adiabatically added flux is shown in Fig.~\ref{fig:ES_flow}(a) for $J_2 = 0,J_\chi/J_1 = 1.0$, where the eigenvalues $- \ln \lambda_i$ in the $S^z = -1, 0, 1$ sectors are demonstrated. 
By analyzing the distributions of the eigenvalues at each given flux, one can find that while the spectrum is symmetric about $S^z = 0$ at zero flux $\theta = 0$, it becomes symmetric about $S^z = 1/2$ at $\theta = 2\pi$ flux, which agrees with a spin-$1/2$ spinon at each end of the cylinder for the system with $\theta = 2\pi$.
Correspondingly, the system evolves from the vacuum sector ($\theta = 0$) to the spinon sector ($\theta = 2\pi$).
With further increase of the flux from $2\pi$ to $4\pi$, the entanglement spectrum becomes symmetric about $S^z = 1$, indicating that the system evolves back to the vacuum sector at $\theta = 4\pi$.

By means of this adiabatic flux insertion, we can also compute the spin Chern number.
In fractional quantum Hall states, a quantized net charge transfer would appear as $\Delta N = C$ from one edge of the sample to the other edge after inserting a period of flux $\theta = 0 \rightarrow 2\pi$, corresponding to a fractionally quantized topological invariant Chern number $C$~\cite{Hatsugai1993}. 
Following this method, we can measure the local spin magnetization moment $\langle S^z_{i,j} \rangle$ for each site $(i,j)$ at each flux~\cite{Gong_2014}.
In the process of inserting flux, one can find that the magnetizations in the bulk remain unchanged but only change near the open boundaries, which is equivalent to spin transfer from one edge to the other one. 
In a period of flux insertion, the transferred total spin gives the spin Chern number.
As shown in Fig.~\ref{fig:ES_flow}(b) for $J_2 = 0$ and $J_\chi/J_1 = 1.0$, the fractionally quantized spin transfer $0.5$ characterizes the spin Chern number $C = 1/2$, which agrees with a Kalmeyer-Laughlin CSL state.
An additional example at $J_2 > 0$ is demonstrated in Fig.~\ref{appfig:ES_flow_J2_03_Jx_07} (see Appendix~\ref{Appendix C:More data of ES and adiabatic flux insertion results}).

\begin{figure}[h]
\includegraphics[width = 0.85\linewidth]{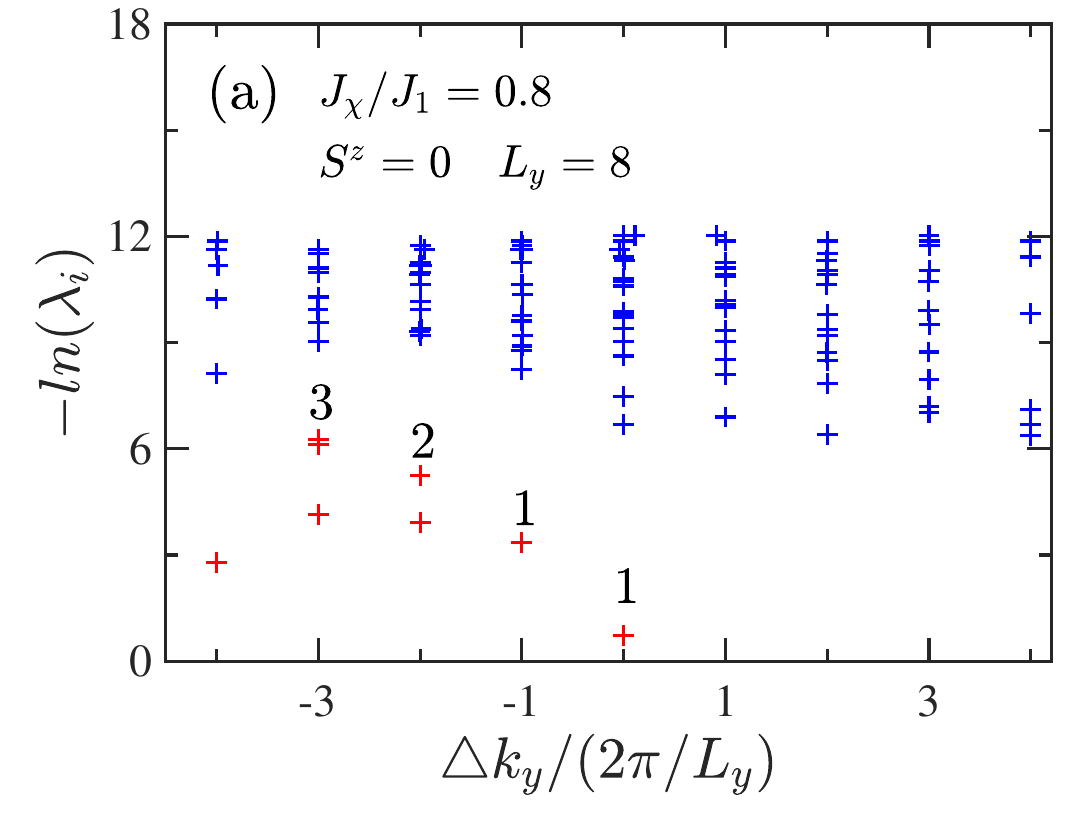}
\includegraphics[width = 0.85\linewidth]{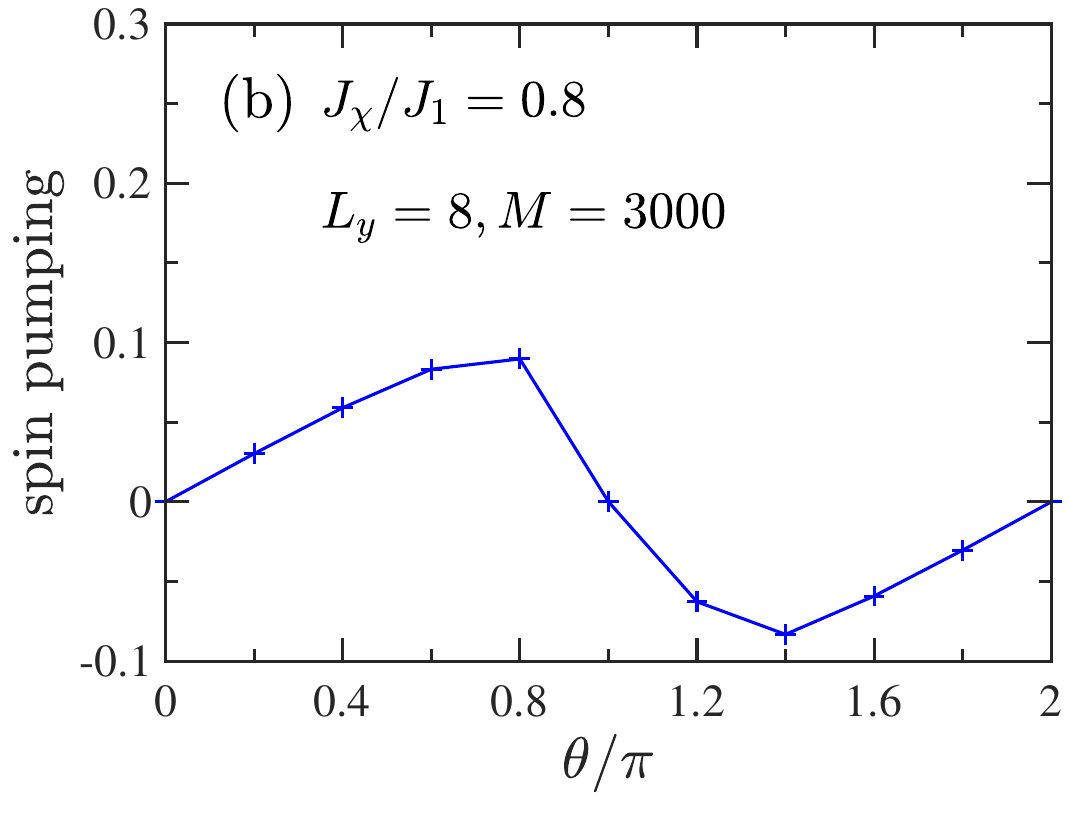}
\caption{Exploring the possible coexistence of magnetic order and topological order.
(a) Entanglement spectrum labeled by the quantum numbers of total spin $S^z = 0$ and relative momentum along the $y$ direction, $\triangle{k_y}$. $\lambda_i$ is the eigenvalue of reduced density matrix. (b) Spin pumping with adiabatically inserted flux $\theta$. We consider the parameter at $J_2/J_1 = 0$, $J_\chi / J_1 = 0.8$ on the $L_y=8$ cylinder.}
\label{appfig:phasecoexist}
\end{figure}

\subsection{Investigation on the possible phase coexistence}

Between the N\'eel AFM and CSL phase, a possible coexistence regime has been proposed based on the mean-field theory~\cite{samajdar2019enhanced}, which is expected to exhibit both the N\'eel AFM order and topological order.
In a recent DMRG study of the spin-$1$ square-lattice $J_1$-$J_2$-$J_{\chi}$ model, a coexistence of stripe AFM order and non-Abelian topological order is suggested, which is characterized by the finite magnetic order and chiral entanglement spectrum~\cite{huang2021}.
Such a combination of numerical results could be taken as the primary evidence of this kind of coexistence of magnetic order and topological order.
In our study, we follow the same strategy to investigate the possible phase coexistence.

We focus on the small parameter regime below the N\'eel order boundary in Fig.~\ref{fig:model}, in which the system still possesses the N\'eel AFM order.
Here, we investigate the possible topological order by studying the entanglement spectrum and Chern number. 
By DMRG simulations, we find that even if we choose the parameter point very close to the boundary such as $J_2 = 0$, $J_{\chi} / J_1 = 0.8$, either the quasidegenerate chiral pattern of the entanglement spectrum or the quantized Chern number is not obtained, as shown in Fig.~\ref{appfig:phasecoexist}.
The entanglement spectrum we obtain should be interpreted as evolving towards that of the CSL with increased chiral coupling. 
On the other hand, above the boundary when the N\'eel AFM order vanishes, both the entanglement spectrum and Chern number simulation results are consistent with an emergent CSL state. 
Near this phase boundary at finite $J_2$, our DMRG results lead to the same conclusion.
Therefore, our results do not support the coexistence of N\'eel order and CSL that was conjectured by mean-field calculations~\cite{samajdar2019enhanced}.

\section{Stripe phase and the magnetic disorder regime}
\label{sec:Phase diagram 2}

For the intermediate $J_2/J_1 \sim 0.55$ regime in the $J_1$-$J_2$ model, we find that a small $J_\chi$ coupling can lead to the emergent CSL state with the characteristic Chern number $C = 1/2$, as shown in Fig.~\ref{appfig:ES_K_flow_J2_05_Jx} of Appendix \ref{Appendix C:More data of ES and adiabatic flux insertion results}.
For the stripe AFM phase at larger $J_2/J_1$, our DMRG results identify either a CSL state or a magnetic disorder state while the stripe order is suppressed by the increased $J_\chi$ coupling. 
In this magnetic disorder state, the entanglement spectrum of the ground state shows the quasidegenerate pattern $\{ 1, 1, 2, 3,...\}$, but the flux insertion simulations fail to obtain a quantized spin Chern number.
Next, we will demonstrate the details of the DMRG results.

\subsection{Stripe phase and the phase transition}

For $J_\chi=0$, the $J_1$-$J_2$ model exhibits a stripe AFM order for $J_2/J_1 \gtrsim 0.61$~\cite{LIU20221034}.
With increased chiral coupling, the stripe order will be suppressed. 
To identify the phase transition with vanishing stripe order, we investigate the spin structure factor and chiral order parameter. 

\begin{figure}[h]
\includegraphics[width = 0.495\linewidth]{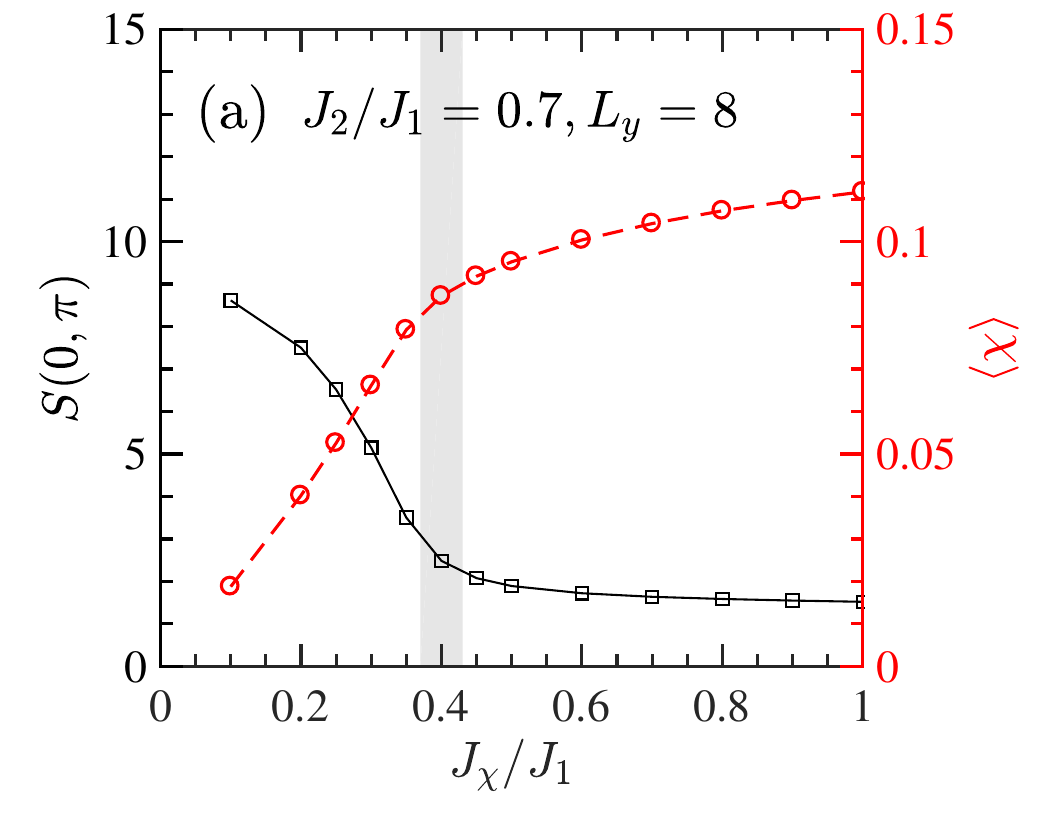}
\includegraphics[width = 0.495\linewidth]{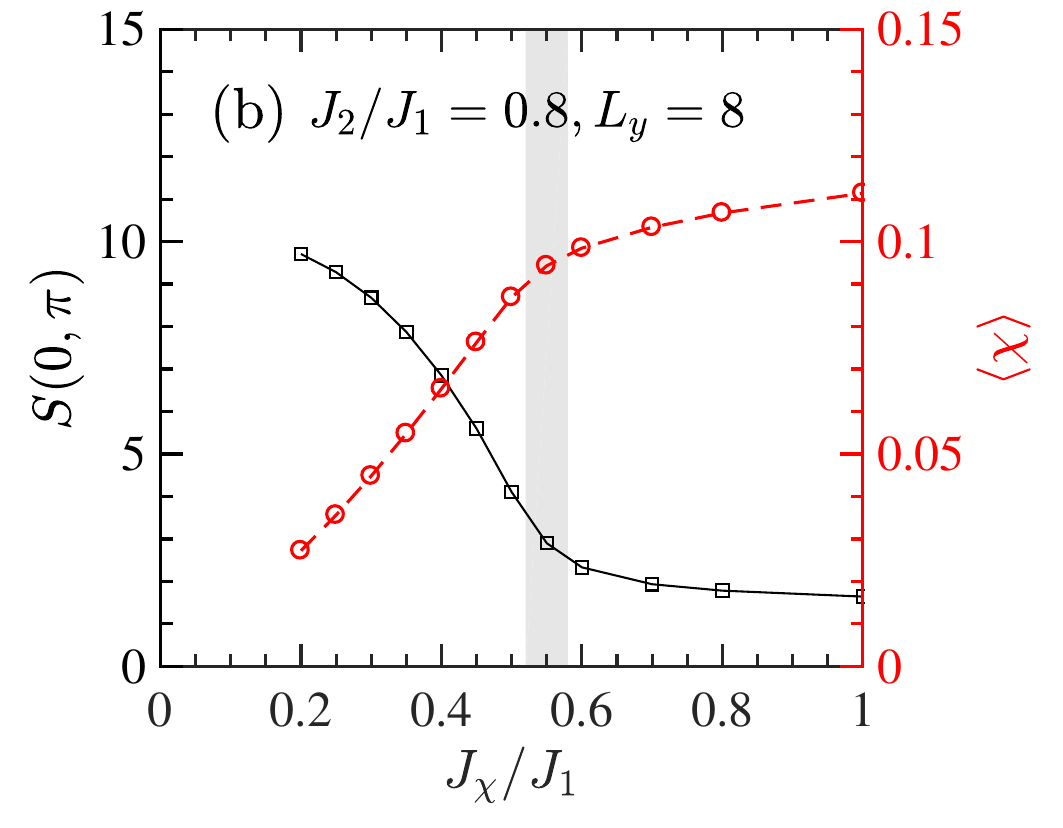}
\caption{Quantum phase transition with vanishing stripe AFM order by tuning chiral coupling. The spin structure factor $S(0,\pi)$ which characterizes the stripe order and the scalar chiral order parameter $\langle \chi \rangle$ are shown versus $J_\chi / J_1$ on the $L_y = 8$ systems for (a) $J_2 = 0.7$ and (b) $J_2 = 0.8$. The shades denote the phase transitions.}
\label{fig:stripe}
\end{figure}

For the stripe AFM order on the square lattice, the spin structure factor on a finite symmetric lattice should have peaks at both $\vec{k} = (0, \pi)$ and $(\pi, 0)$.
In our DMRG calculations on long-cylinder geometry, the spin correlations form the stripe pattern along a certain direction because of the geometry, and the spin structure factor selects the peak at $\vec{k} = (0, \pi)$.
Therefore, we study $S(0, \pi)$ versus the tuning chiral coupling.
In Fig.~\ref{fig:stripe}, we show the structure factor $S(0,\pi)$ and chiral order parameter $\langle \chi \rangle$ for $J_2/J_1 = 0.7$ and $0.8$ on the $L_y = 8$ cylinder.
As marked by the shades, both quantities consistently characterize a phase transition with the vanished stripe AFM order.
One can find that while the $J_{\chi}$ dependence of chiral order exhibits a clear change characterizing a continuous-like phase transition (the red curve), the stripe AFM structure factor $S(0,\pi)$ drops fast to quite small values (the black curve), which is consistent with the transition described by the change of chiral order $\langle \chi \rangle$.

In the spin-$1$ $J_1$-$J_2$-$J_\chi$ model, the DMRG calculations find evidence for the coexistence of stripe order and non-Abelian topological order~\cite{huang2021}.
Therefore, in the stripe AFM phase of the studied spin-$1/2$ model, we also examine the entanglement spectrum and Chern number, but our results do not support a coexistent topological order in the stripe phase (not shown here).

\subsection{The magnetic disorder regime}

\begin{figure}[th]
\includegraphics[width = 0.85\linewidth]{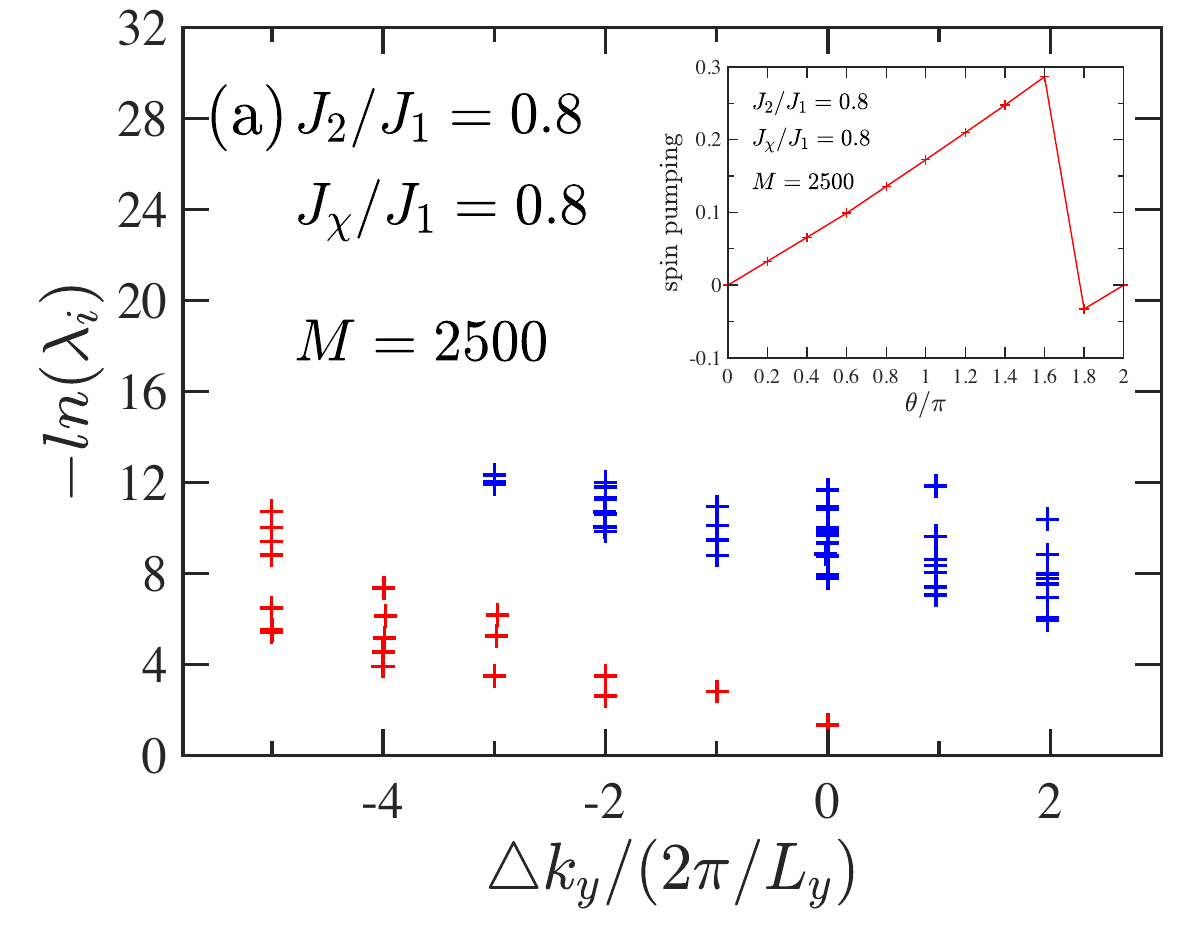}
\includegraphics[width = 0.85\linewidth]{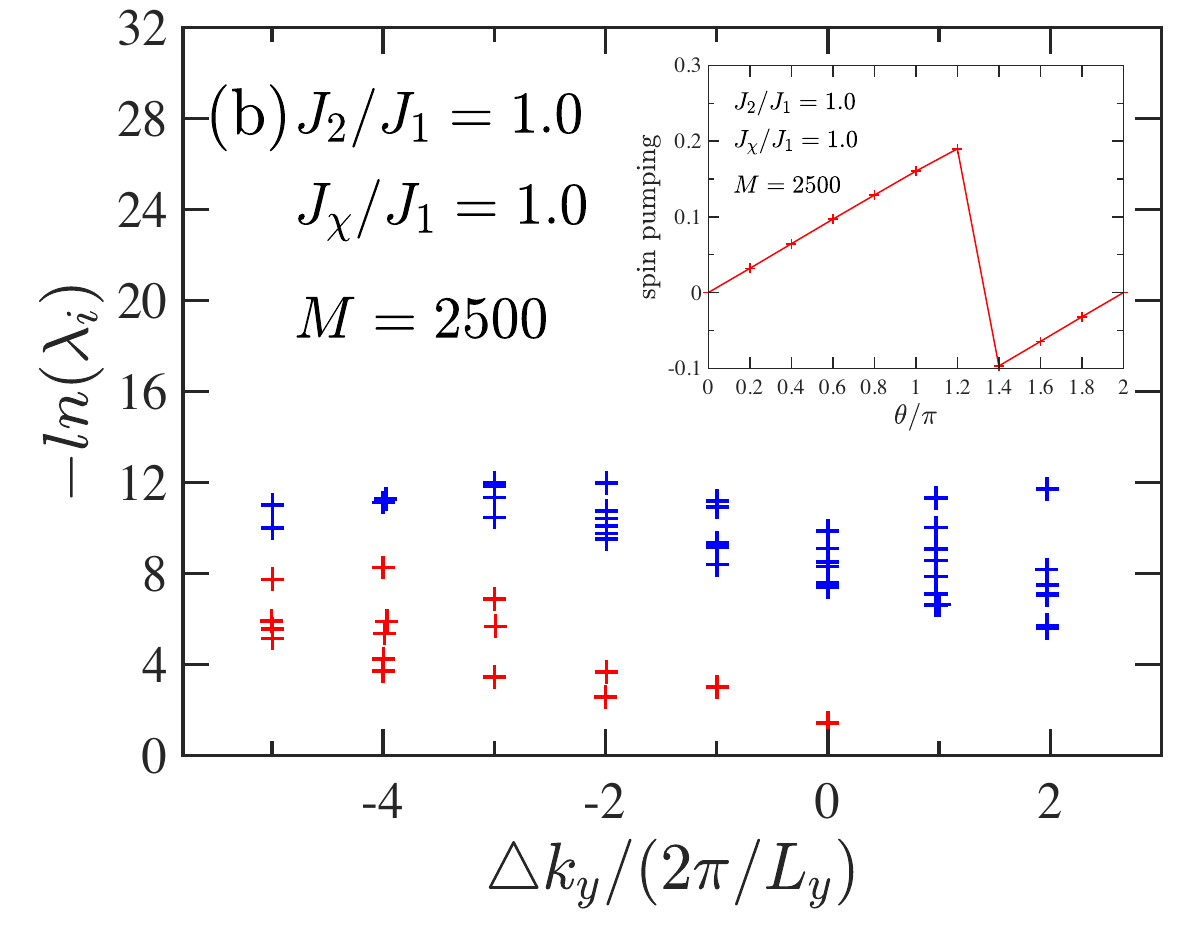}
\caption{Entanglement spectrum and adiabatic flux insertion simulations in the magnetic disorder regime. (a) and (b) show the entanglement spectrum labeled by the quantum numbers of total spin $S^z = 0$ and relative momentum along the $y$ direction $\triangle{k_y}$ for the ground states of $J_2 / J_1 = J_{\chi} / J_1 = 0.8$ and $J_2 / J_1 = J_{\chi} / J_1 = 1.0$, respectively, on the $L_y = 8$ cylinder systems. $\lambda_i$ are the eigenvalues of reduced density matrix. The results are obtained by keeping $2500$ U(1) states. The insets show the spin pumping with adiabatically inserted flux $\theta$. At a flux smaller than $2\pi$, the spin accumulation collapses. As a result, the obtained ground state at $\theta = 2\pi$ is the same as that at zero flux.}
\label{fig:ES_K_flow_J2_Jx}
\end{figure}

Next, we focus on the discussion of the magnetic disorder regime based on the DMRG results.
In this magnetic disorder regime, the entanglement spectrum of the obtained ground state also exhibits the quasidegenerate chiral counting $\{ 1, 1, 2, 3, 5,...\}$ and a large entanglement gap that separates the quasidegenerate levels from the higher levels, as shown in Fig.~\ref{fig:ES_K_flow_J2_Jx} of the $L_y = 8$ systems. 
The entanglement spectrum results may indicate the obtained ground state as the vacuum sector of a CSL state.
Nonetheless, the spin accumulation in the adiabatic flux insertion simulation is found to collapse at a flux $\theta$ smaller than $2\pi$, as demonstrated in the insets of Fig.~\ref{fig:ES_K_flow_J2_Jx}.
Consequently, the ground state at $\theta = 2\pi$ is the same as that at $\theta = 0$, namely, the flux insertion simulations fail to find the spinon topological sector as expected for a CSL.

Furthermore, we explore possible magnetic order in this regime and provide DMRG data to support the absent magnetic order. 
First of all, we carefully examine the spin structure factor in the region of $0.6 \leq J_2/J_1 \leq 1.0$ and $0 \leq J_{\chi}/J_1 \leq 1.5$ on the $L_y = 8$ cylinder.
We find that when the stripe AFM order is suppressed as we identify in Fig.~\ref{fig:stripe}, the spin structure factor shows small peaks at $\vec{k} = (0, \pi)$, $(\pi, 0)$, and $(\pi/2, \pi/2)$ in the magnetic disorder regime (see Fig.~\ref{fig:CSS_Sq} in Appendix \ref{Appendix E: Analyses of magnetic order parameters}), which implies a possible weak CSS magnetic order~\cite{rabson1995}.
To detect such a possible magnetic order, we calculate the corresponding magnetic order parameters $m^2(0,\pi)$, $m^2(\pi,0)$ and $m^2(\pi/2,\pi/2)$ and make the finite-size extrapolations to estimate the results in the thermodynamic limit. 
Notice that we only consider the results on the $L_y = 4, 8, 12$ systems, which are compatible with the structure factor peak at $\vec{k} = (\pi/2, \pi/2)$.

\begin{figure}[h]
\includegraphics[width = 0.85\linewidth]{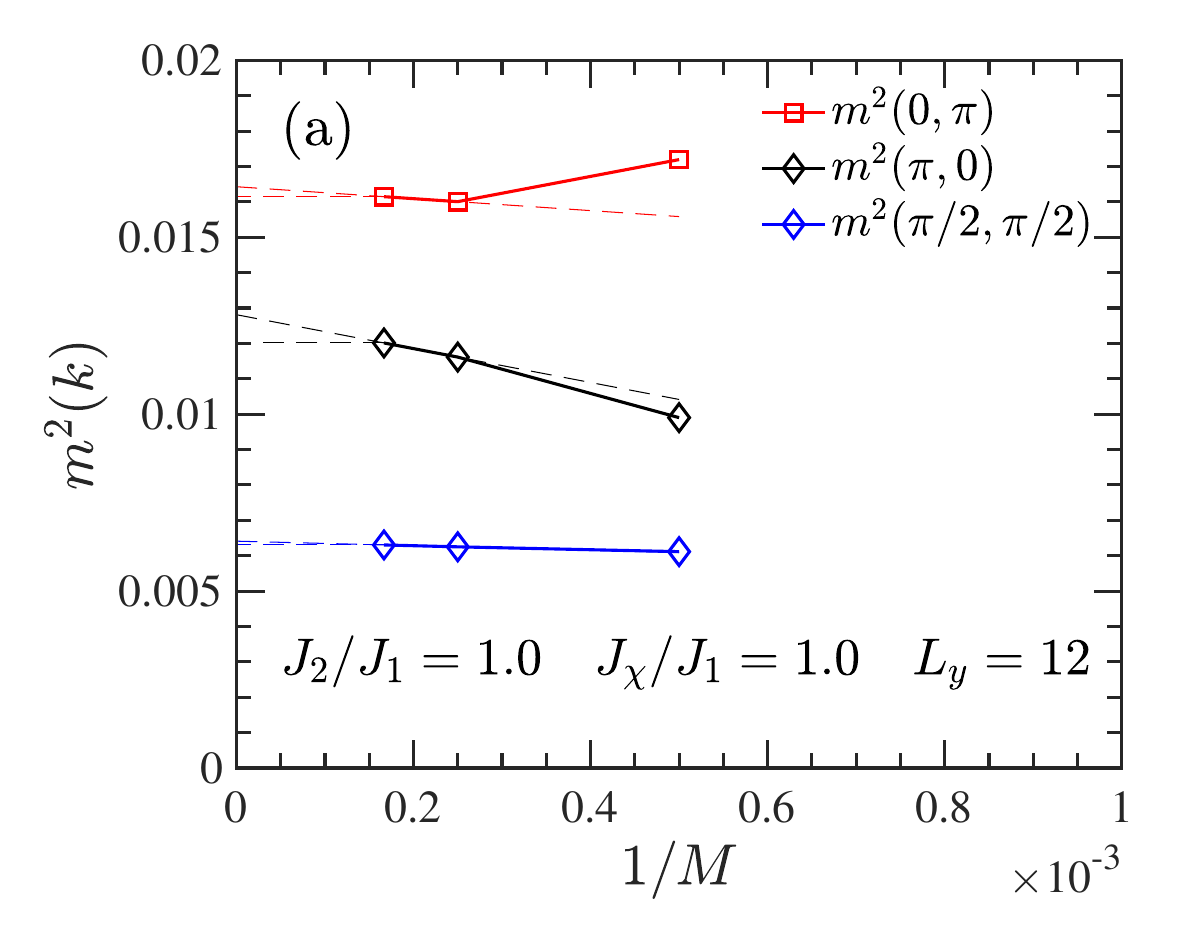}
\includegraphics[width = 0.85\linewidth]{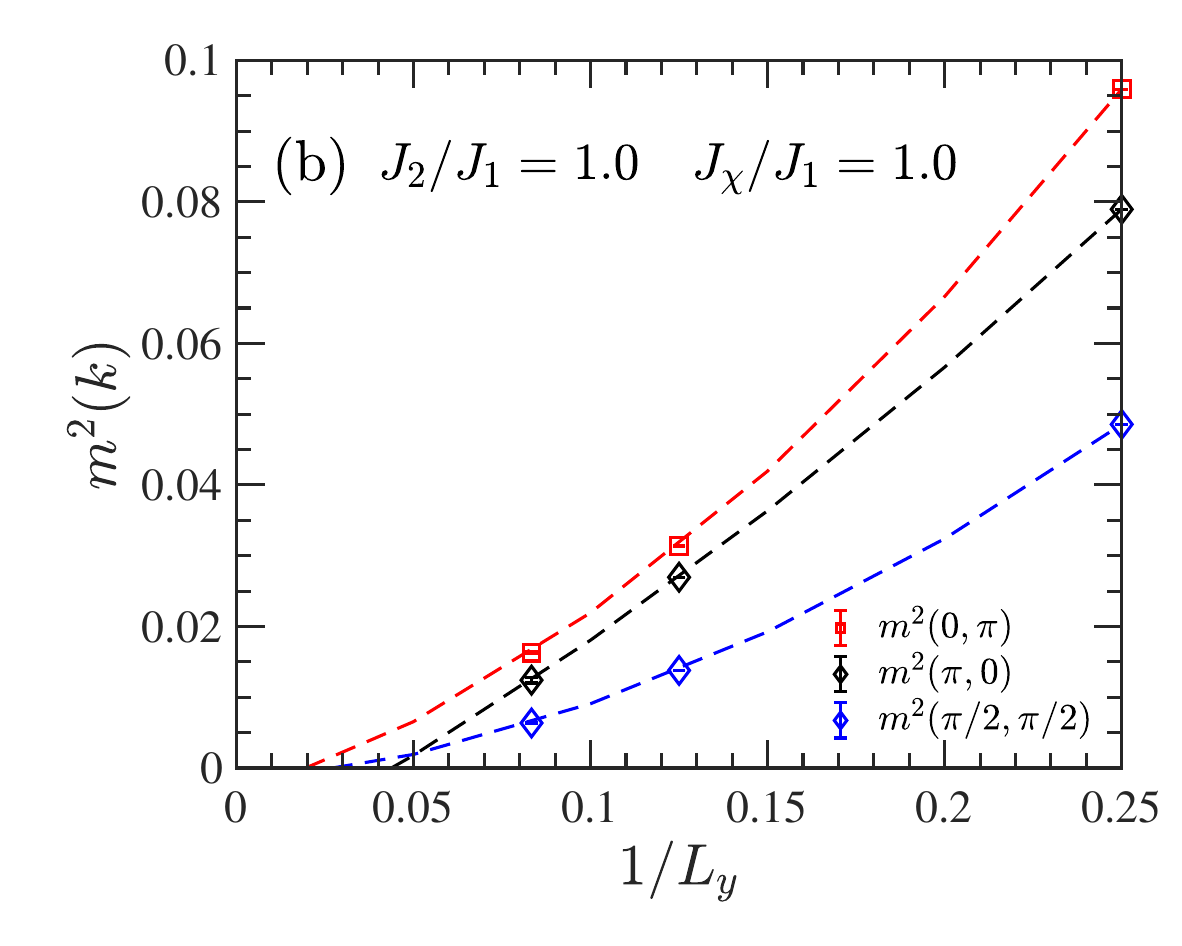}
\caption{Analyses of magnetic order parameters at $J_2/J_1 = J_{\chi}/J_1 = 1.0$. (a) Bond dimension dependence of the magnetic order parameters $m^2(0, \pi)$, $m^2(\pi, 0)$, and $m^2(\pi/2, \pi/2)$ with bond dimensions $M = 2000$, $4000$, and $6000$ on the $L_y = 12$ cylinder. We estimate the error bar of the extrapolation by using the $M = 6000$ result and the linearly extrapolated result by using $M = 4000$ and $6000$ data (the dashed line). (b) Finite-size scaling of the magnetic order parameters $m^2(0, \pi)$, $m^2(\pi, 0)$, and $m^2(\pi/2, \pi/2)$ with $L_y = 4, 8, 12$. The error bars of the $L_y = 12$ data determined from (a) are also shown, which are very small.}
\label{fig: CSS_m2}
\end{figure}

We first examine the coupling dependence of the spin structure factor at $\vec{k} = (0, \pi)$, $(\pi, 0)$, and $(\pi/2, \pi/2)$ on the $L_y = 8$ system (see Fig.~\ref{fig:CSS_Sq} in Appendix \ref{Appendix E: Analyses of magnetic order parameters}), which shows us the parameter points with the strongest structure factors on this finite-size system.
Furthermore, we make the finite-size scaling of these magnetic order parameters at the parameter points with strong spin structure factor.
Due to the computation cost of the complex wavefunction, we keep the bond dimensions up to $6000$ SU(2) multiplets, which can give well-converged results for $L_y = 4$ and $8$, but less converged results for $L_y = 12$.
Therefore, we analyze the bond dimension dependence of the magnetic order parameters for $L_y = 12$ to estimate the results in the infinite-bond-dimension limit, as shown in Fig.~\ref{fig: CSS_m2}(a) for $J_2/J_1 = J_{\chi}/J_1 = 1.0$.
We keep $M = 2000$, $4000$, and $6000$ SU(2) multiplets to obtain the magnetic order parameters, and plot the results versus $1/M$.
While $m^2(0,\pi)$ and $m^2(\pi/2,\pi/2)$ seem to approach convergence with bond dimension, the extrapolated $m^2(\pi,0)$ may have the relatively larger error bar. 
We estimate the error bar of the extrapolation by using the $M = 6000$ result and the linearly extrapolated result by considering $M = 4000$ and $6000$, which are taken as the lower and upper boundaries of the $M \rightarrow \infty$ result.
In Fig.~\ref{fig: CSS_m2}(b), we make the finite-size scaling for the three order parameters with $L_y = 4$, $8$, $12$, which all decay fast with system width and are smoothly extrapolated to vanish.
Notice that since the error bars of the $L_y = 12$ results are quite small, the uncertainties of the results do not change our conclusion.
We have also checked the size-scaling results at $J_2/J_1=0.8$, $J_\chi/J_1=0.7$ and $J_2/J_1=0.8$, $J_\chi/J_1=1.0$ (see Fig.~\ref{fig:CSS_J2_08} in Appendix \ref{Appendix E: Analyses of magnetic order parameters}), which all agree with no magnetic order.
Therefore, our finite-size scaling analyses of the magnetic order parameters suggest no magnetic ordering in this regime, which could be consistent with the ground state that appears like a CSL state in the vacuum topological sector, as characterized by the entanglement spectrum in Fig.~\ref{fig:ES_K_flow_J2_Jx}.

\begin{figure}[ht]
\includegraphics[width = 0.85\linewidth]{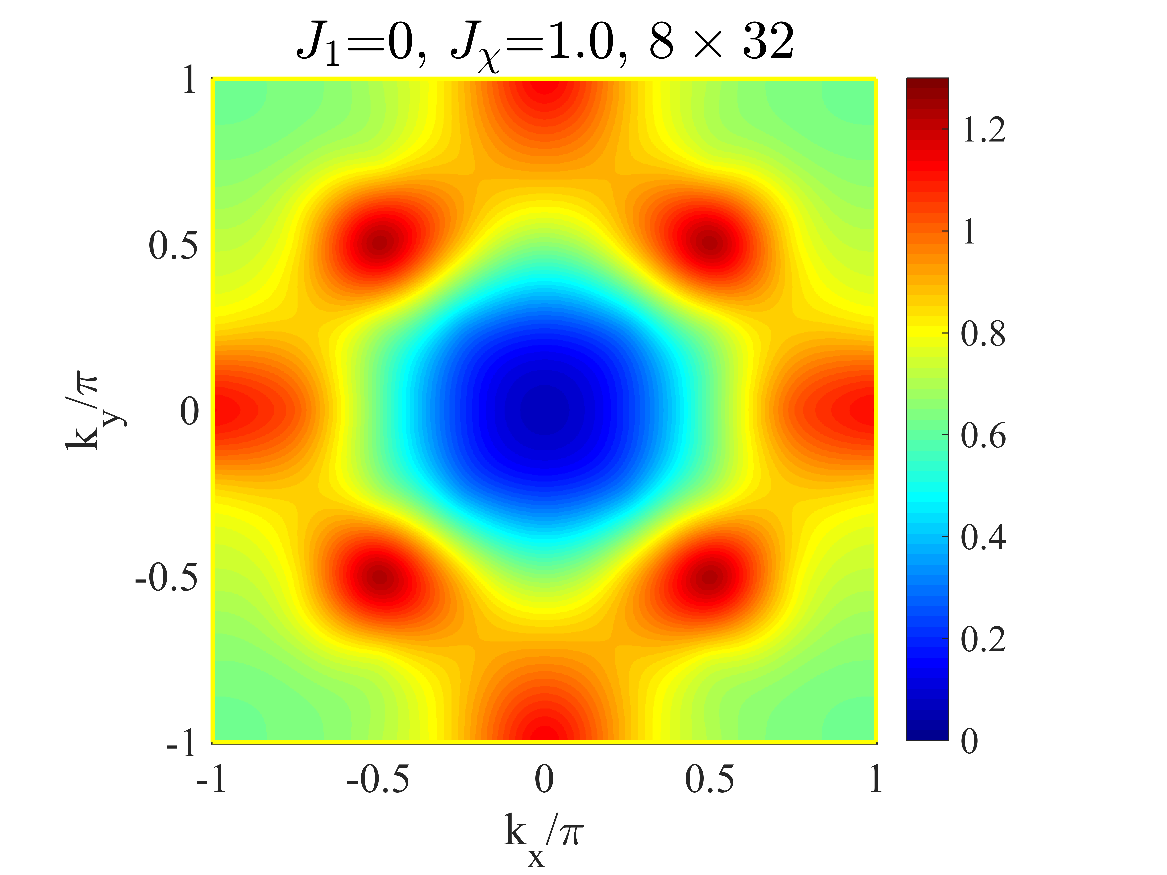}
\caption{Spin structure factor for $J_1 = 0$, $J_{\chi} = 1.0$.
The structure factor is obtained from the Fourier transformation of spin correlations of the middle $L_y \times L_y$ sites on the $L_y=8, L_x = 32$ cylinder, which shows round peaks at ${\bf k} = (\pi, 0)$, $(0, \pi)$, and $(\pm \frac{\pi}{2}, \pm \frac{\pi}{2})$. These peak momenta agree with that in the CSS magnetic order state, but the very broad peaks indicate the vanished magnetic order.
}
\label{fig:CSS_sq}
\end{figure}

\section{Strong tendency to form a dimer order at large chiral coupling}
\label{sec:VBC}

In this section, we extend our discussion to the large chiral coupling regime, which is beyond the parameter region of Fig.~\ref{fig:model}. 
Classically, the dominant chiral coupling $J_{\chi}$ in this system will lead to the magnetically ordered CSS~\cite{rabson1995}.
In a recent DMRG study of the spin-$1$ $J_1$-$J_2$-$J_{\chi}$ model, the CSS has also been identified in the large-$J_{\chi}$ regime~\cite{huang2021}.

Here, we examine the spin-$1/2$ case.
We first measure the spin structure factor to detect the possible CSS magnetic order. 
Since the spin pattern of the CSS order has the translation period of $4$ along both the $x$ and $y$ directions (see Fig.~\ref{fig:phase diagram_J1_Jx} in Appendix~\ref{Appendix A:Classical phase diagram}), we choose the system circumference $L_y = 4, 8$ to accommodate this spin pattern.
Indeed, the spin structure factor $S(\bf{k})$ exhibits peaks at ${\bf k} = (\pi, 0), (0, \pi)$ and $(\pm \frac{\pi}{2}, \pm \frac{\pi}{2})$, as shown in Fig.~\ref{fig:CSS_sq} of the pure $J_{\chi}$ model on the $L_y = 8$ system.
The peak momenta fully agree with the feature of the CSS state~\cite{huang2021}.
Nonetheless, one can find in Fig.~\ref{fig:CSS_sq} that all the peaks of $S(\bf{k})$ are very broad, which indicates that the CSS magnetic order is likely to be melted by the stronger quantum fluctuations in the spin-$1/2$ case while this order can persist in the spin-$1$ case~\cite{huang2021}.
One possibility of this nonmagnetic state could be the CSL, which however is not supported by the DMRG results.

\begin{figure}[h]
\includegraphics[width = 0.85\linewidth]{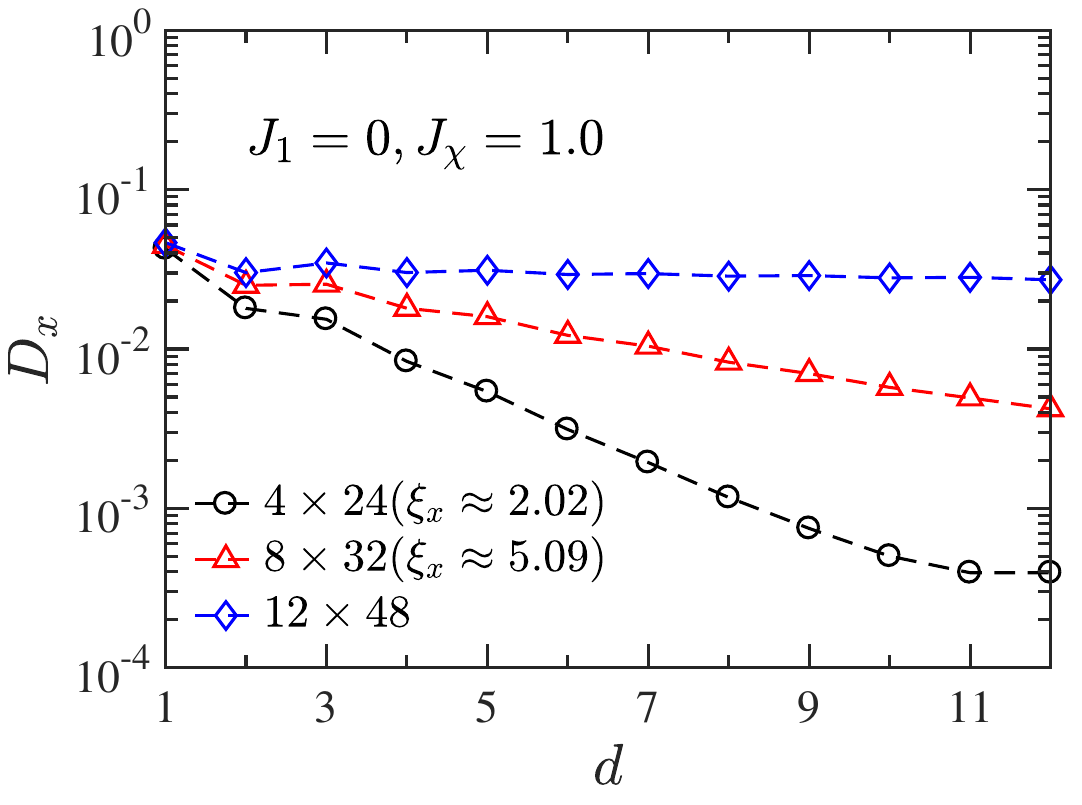}
\caption{Log-linear plot of the horizontal bond dimer order $D_x$ versus the distance $d$ away from the open boundary. The system is the pure $J_\chi$ model with $J_\chi = 1.0$, $J_1 = J_2 = 0$ on the $L_y = 4, L_x = 24$, $L_y = 8, L_x = 32$, and $L_y = 12, L_x = 48$ cylinders. The exponential fitting of $D_x \sim e^{-d/\xi_{x}}$ gives the decay length $\xi_x \approx 2.02$ and $5.09$ for the $L_y = 4$ and $8$ cylinders, respectively. The $L_y = 12$ results are obtained with the bond dimensions $M = 5000$.
}
\label{fig:bond_dimer}
\end{figure}

Therefore, we turn to explore a possible translational symmetry breaking for large-$J_{\chi}$ coupling.
To accommodate the short-range spin configurations as shown in Fig.~\ref{fig:CSS_sq}, we study the $L_y = 4, 8, 12$ cylinders.
However, at large $J_{\chi}$, the DMRG calculations are harder to converge, which makes the good convergence for $L_y = 12$ more difficult.
Thus, we will mainly rely on the well-converged $L_y = 4, 8$ results and also refer to the less-converged results for $L_y = 12$.

We study the decay length of the boundary-induced dimer order and its circumference dependence, which has been found efficient to identify a bond dimer state even if the dimer order is very weak~\cite{sandvik2012,gong2014}.
Here, we briefly introduce the strategy of this method. 
Since our DMRG calculations are performed on the cylinder geometry with open boundary conditions, the translational symmetry along the $x$ direction has already been broken, which leads to a dimer order of the bond energy $\langle {\bf S}_i \cdot {\bf S}_j \rangle$.
We define the dimer order parameter $D_x$ as the bond energy difference between the two neighboring horizontal bonds
\begin{equation}
D_{x}(d) = \langle {\bf S}_{(d, y)} \cdot {\bf S}_{(d+1, y)} \rangle - \langle {\bf S}_{(d+1, y)} \cdot {\bf S}_{(d+2, y)} \rangle,
\end{equation}
where ${\bf S}_{(d,y)}$ denotes the spin on the position $(d,y)$. 
Therefore, on finite-size systems $D_{x}$ would decay exponentially from the edge to the bulk as $D_{x} \sim e^{-d/\xi_x}$.
If the ground state is a dimer order state in two dimensions, the decay length $\xi_x$ would grow rapidly with increased system circumference and eventually diverge in a certain width when spontaneous symmetry breaking can happen.
Otherwise, in a state without a dimer order, the decay length $\xi_x$ would be a finite number in the thermodynamic limit.

Following this strategy, we compute the dimer order $D_{x}$ on different systems.
In Fig.~\ref{fig:bond_dimer}, we show the log-linear plot of $D_{x}$ for $J_\chi = 1.0$, $J_1 = 0$ on different sizes.  
On the smaller $L_y = 4, 8$ systems, $D_{x}$ decays exponentially and $\xi_x$ grows rapidly from $\xi_x \simeq 2.02$ to $5.09$.
On the wider $L_y = 12$ system, we push the calculations up to $5000$ SU(2) multiplets, and the results show a strong dimer order.
Although the $L_y = 12$ results are less converged, the strong dimer order and the quickly growing $\xi_x$ from $L_y = 4$ to $L_y = 8$ strongly suggest a dimer order in two-dimensional limit. 
This tendency to develop a dimer order at large chiral coupling is also suggested by the growing $\xi_x$ with increased $J_{\chi}$ coupling (see Fig.~\ref{fig:bond_dimer_J2_05} in Appendix ~\ref{Appendix F: VBC phase}).
While the CSS is robust in the spin-$1$ system, the magnetic order seems to give way to a magnetically disordered dimer state in the spin-$1/2$ system due to the stronger quantum fluctuations.

\section{Summary and discussion}
\label{sec:Summary and discussion}

We have studied the quantum phase diagram of the spin-$1/2$ $J_1$-$J_2$-$J_\chi$ model on the square lattice using DMRG calculations.
This model was expected to give some insight for the giant thermal Hall conductivity in cuprate superconductors~\cite{samajdar2019enhanced,samajdar2019thermal,Zhang2021,merino2022majorana,Han2019,li2019thermal,li2019theory}.
By mean-field calculation, a phase coexistence of N\'eel AFM order and topological CSL was proposed in this model, which could be used to explain the enhanced thermal Hall response in some cuprate materials~\cite{samajdar2019enhanced,Zhang2021}. 
On the other hand, a recent DMRG study of the spin-$1$ $J_1$-$J_2$-$J_\chi$ model identifies the coexistence of stripe AFM order and non-Abelian CSL topological order.
The reported coexistence of magnetic order and topological order motivates us to examine the quantum phases and possible coexistence in this spin-$1/2$ model.

By tuning the interactions $0 \leq J_2/J_1 \leq 1.0$ and $0 \leq J_\chi/J_1 \leq 1.5$, we identify the N\'eel AFM phase, stripe AFM phase, CSL phase, and a magnetic disorder regime, as shown in Fig.~\ref{fig:model}. 
For the intermediate nonmagnetic region $J_2 / J_1 \simeq 0.5$ of the $J_1$-$J_2$ model, we find that a small chiral coupling can drive the emergence of the CSL state.
However, the related phase transition requires numerical studies of much larger length scales, which is beyond the scope of this work.

With growing chiral coupling, we identify the phase transition with vanishing N\'eel AFM order by examining the N\'eel order parameter $m^2(\pi,\pi)$ and scalar chiral order parameter $\langle \chi \rangle$. 
Above this phase boundary, we can verify the Kalmeyer-Laughlin CSL phase by the characteristic features including the quantized topological entanglement entropy $\gamma = (\ln 2)/2$, chiral entanglement spectrum agreeing with the SU(2)$_1$ conformal field theory, and spin Chern number $C = 1/2$.
Below the phase boundary when the N\'eel order exists, we explore the possible phase coexistence. 
Nonetheless, either the entanglement spectrum or Chern number result does not support a coexistent topological order.
Even near the phase boundary with a very weak N\'eel order, the entanglement spectrum does not form the chiral structure of the CSL but exhibits an evolution towards the spectrum of a CSL, which suggests no coexistence of magnetic and topological order.

For the stripe phase at larger $J_2/J_1$, we find two regimes when the stripe AFM order is suppressed by increased chiral coupling.
One regime is identified as the same CSL phase, and the other one is a magnetic disorder regime.
In the stripe and CSL phase, our DMRG results do not support coexistence of magnetic and topological order.
In the magnetic disorder regime, the spin structure factor shows broad peaks at ${\bf k} = (0, \pi)$, $(\pi, 0)$ and $(\pm \pi/2, \pm \pi/2$), which agrees with the spin configuration of the CSS magnetic order.
However, finite-size scaling of magnetic order parameters suggests no magnetic ordering in the thermodynamic limit. 
On the other hand, the entanglement spectrum of the ground state exhibits the quasidegenerate levels that are consistent with the CSL.
Therefore, the obtained ground state appears like the vacuum-sector ground state of a CSL.
The absent spinon sector in the flux insertion simulation implies that the state corresponding to the ground state in the spinon sector may intersect with higher-energy states.
The exact nature of this magnetic disorder regime needs further studies on larger system size or by investigating other properties.

Beyond the phase diagram in Fig.~\ref{fig:model}, we also study the $J_\chi$-dominant regime. 
While the spin-$1$ $J_1$-$J_2$-$J_\chi$ model still possesses the CSS state in this regime, this magnetic order seems to give way to a disordered dimer state in the studied spin-$1/2$ system due to the stronger quantum fluctuations.
Our results show the striking differences between the spin-1/2 and spin-1 systems, and provide numerical insights for further understanding on the coexistence of conventional order and topological order.
\\

{\it Note added.} Recently, we became aware of an article by Yang {\it et al.}~\cite{yang2024}, who studied the same model by using numerical simulations. Both of the two works consistently find the N\'eel AFM, stripe AFM, chiral spin liquid, and magnetic disorder (called the nematic spin liquid in their paper) phases.
While Ref.~\cite{yang2024} proposes a magnetically ordered chiral spin state based on the enhanced spin structure factor, our finite-size extrapolation results suggest that the chiral spin state is unlikely in the studied parameter regime.

\begin{acknowledgments}
We acknowledge the stimulating discussions with Jian-Wei Yang, Wei Zhu, and Ling Wang.
X.T.Z. and S.S.G. were supported by the National Natural Science Foundation of China (Grants No. 11874078 and No. 11834014) and the Special Project in Key Areas for General Universities in Guangdong Province (Project No. 2023ZDZX3054). 
Y.H. was supported by the U.S. DOE NNSA under Contract No. 89233218CNA000001 and by the Center for Integrated Nanotechnologies, a DOE BES user facility, in partnership with the LANL Institutional Computing Program for computational resources. H.Q.W. was supported
by Guangdong Provincial Key Laboratory of Magnetoelectric Physics and Devices (Grant No. 2022B1212010008) and Guangzhou Basic and Applied Basic Research Foundation (Grant No.202201011569). D.N.S. was supported by the National Science Foundation through the Partnership in Research and Education in Materials Grant No. DMR-1828019. S.S.G. and H.Q.W. also acknowledge the support from the Dongguan Key Laboratory of Artificial Intelligence Design for Advanced Materials. 
\end{acknowledgments}

\appendix

\section{Magnetically ordered chiral spin state}
\label{Appendix A:Classical phase diagram}

\begin{figure}[ht]
\includegraphics[width = 0.7\linewidth]{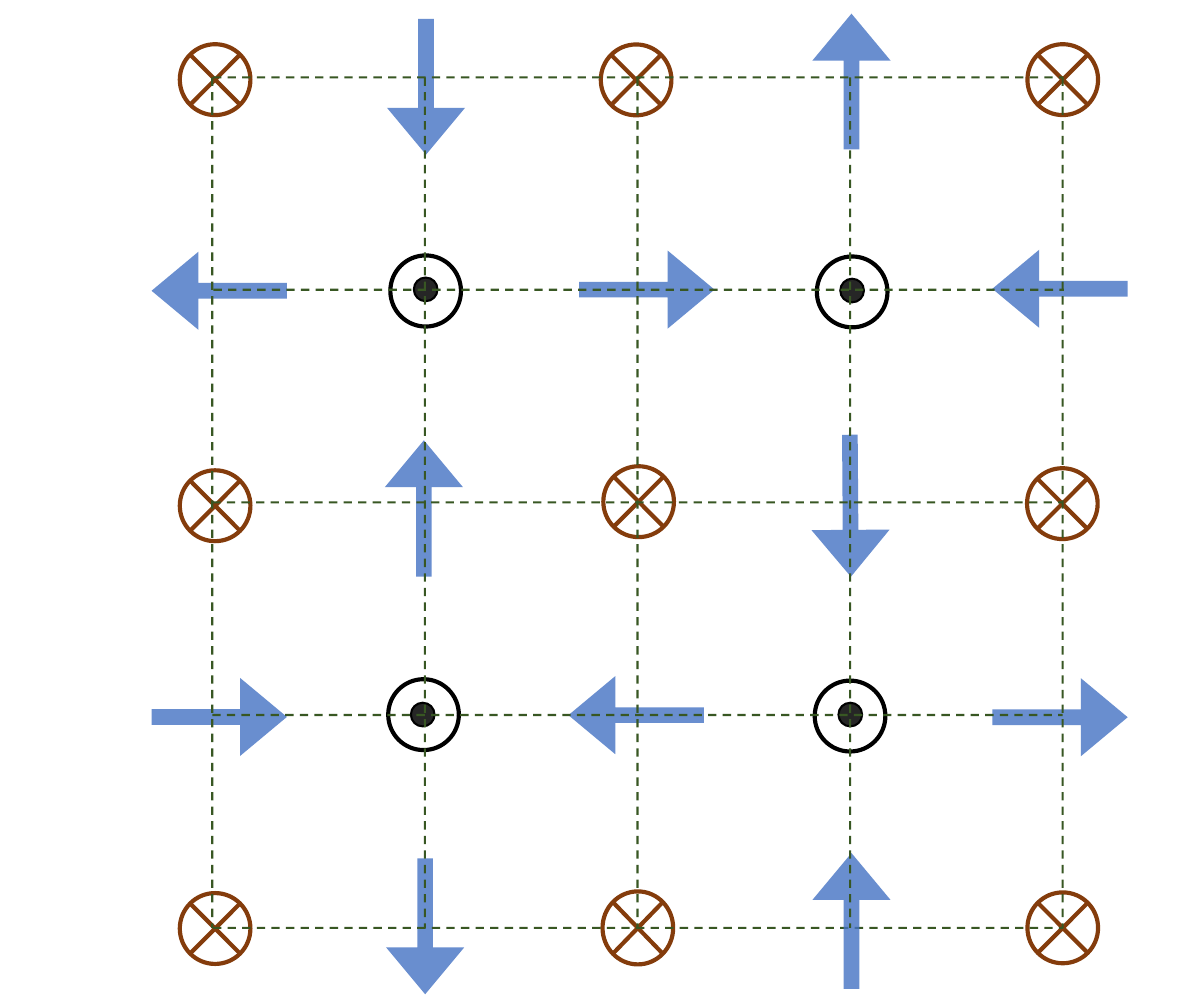}
\caption{Spin configuration of the classical magnetic chiral spin state. The arrows denote the spins pointing to the different directions. 
}
\label{fig:phase diagram_J1_Jx}
\end{figure}

To demonstrate the magnetic chiral spin state (CSS), we plot the spin configuration of the classical CSS in Fig.~\ref{fig:phase diagram_J1_Jx}, where the different arrows denote the spins pointing to the different directions.
One can find that the spin configuration has the translation period of $4$ along both the $x$ and $y$ directions.
Classically, the square-lattice system with only the three-spin scalar chiral interaction $J_{\chi} {\bf S}_i \cdot ({\bf S}_j \times {\bf S}_k)$ can lead to this CSS.

In the quantum case, the spin-$1$ square-lattice $J_1$-$J_2$-$J_\chi$ model has been studied by DMRG calculation~\cite{huang2021}, which focuses on the intermediate $0.4 \leq J_2/J_1 \leq 0.6$. 
In this region, a relatively small chiral coupling $J_\chi/J_1 \gtrsim 0.4$ can induce a robust CSS magnetic order~\cite{huang2021}.

\section{More data about phase transition}
\label{Appendix B:Quantum phase transition}

In Fig.~\ref{fig:CSL} of the main text, we have shown the DMRG results for determining the transition with vanishing the N\'eel AFM order at $J_2 = 0$.
Here, we demonstrate more DMRG data to support the similar transition with increasing $J_\chi$ at finite $J_2$ in Fig.~\ref{appfig:J_nn_01,J_nn_03,J_nn_04}, where the derivative of the bulk chiral order $\langle \chi \rangle$ with respect to $J_\chi/J_1$ clearly characterizes the phase transition.

\begin{figure}[ht]
\includegraphics[width = 0.85\linewidth]{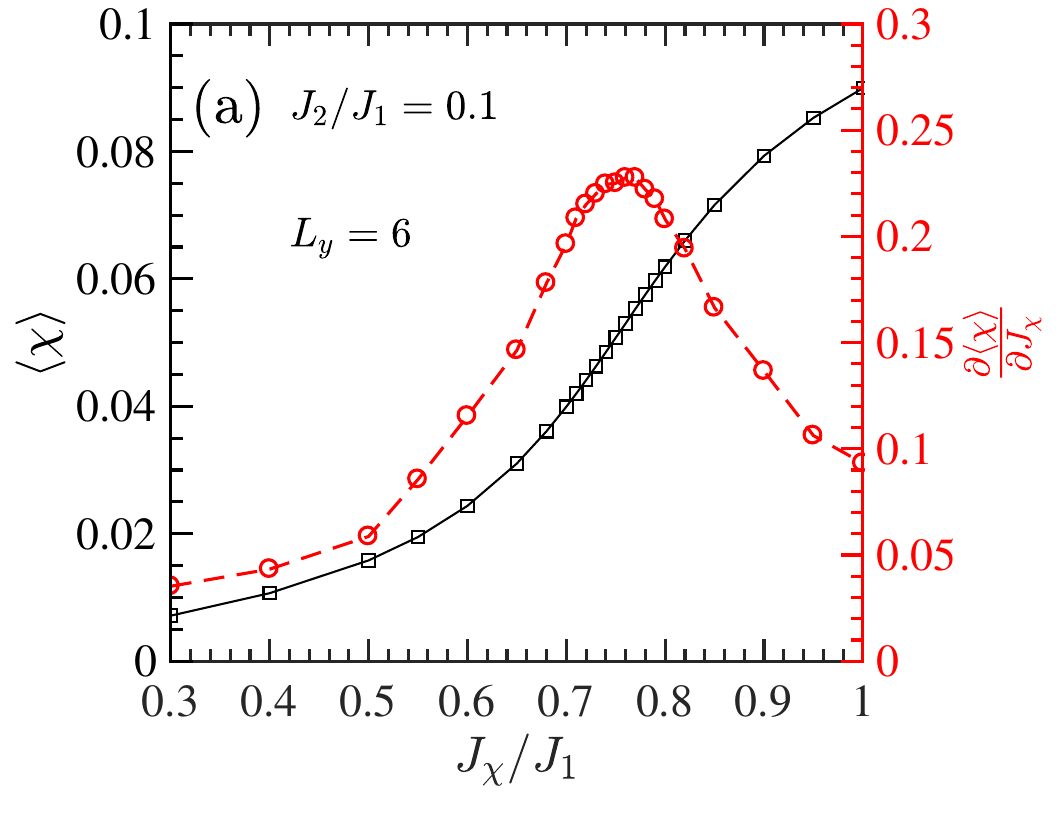}
\includegraphics[width = 0.85\linewidth]{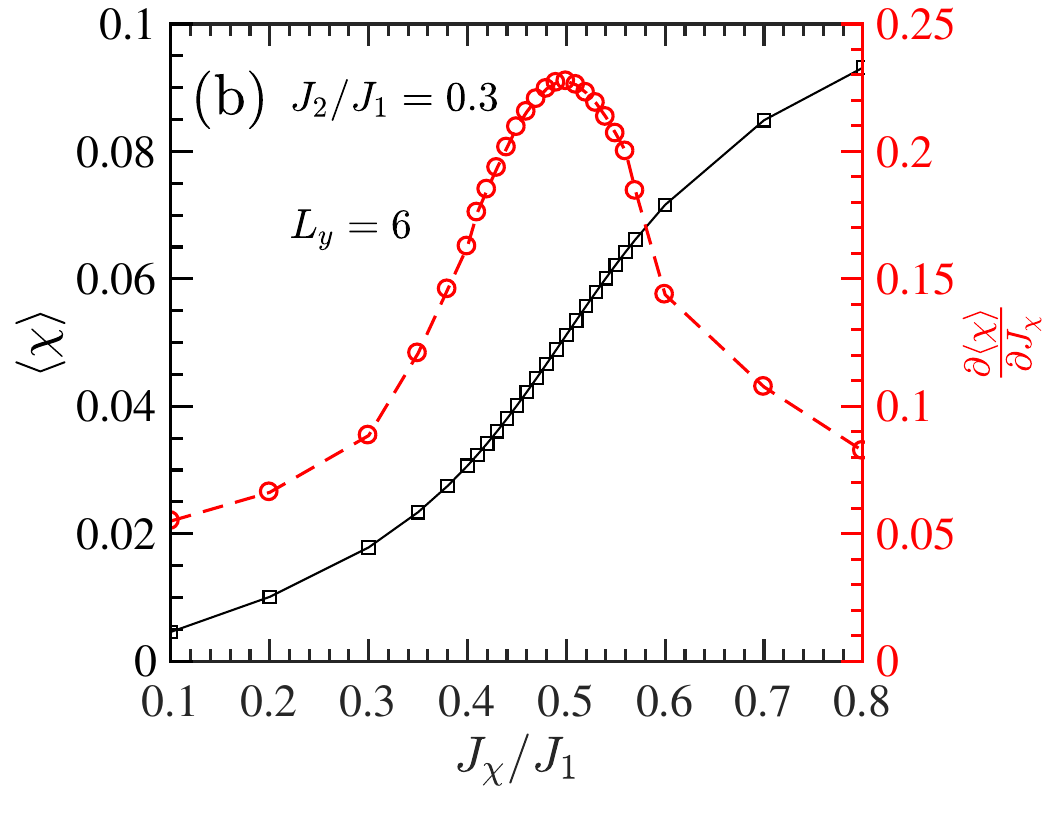}
\caption{The bulk chiral order $\langle \chi \rangle$ versus $J_\chi/J_1$ and its derivative $\partial \langle \chi \rangle / \partial J_\chi$ for $J_2/J_1 = 0.1$ and $0.3$ on the $L_y = 6$ systems.}
\label{appfig:J_nn_01,J_nn_03,J_nn_04}
\end{figure}

\section{More data of entanglement spectrum and adiabatic flux insertion results in the chiral spin liquid phase}
\label{Appendix C:More data of ES and adiabatic flux insertion results}

\begin{figure}[ht]
\includegraphics[width = 0.85\linewidth]{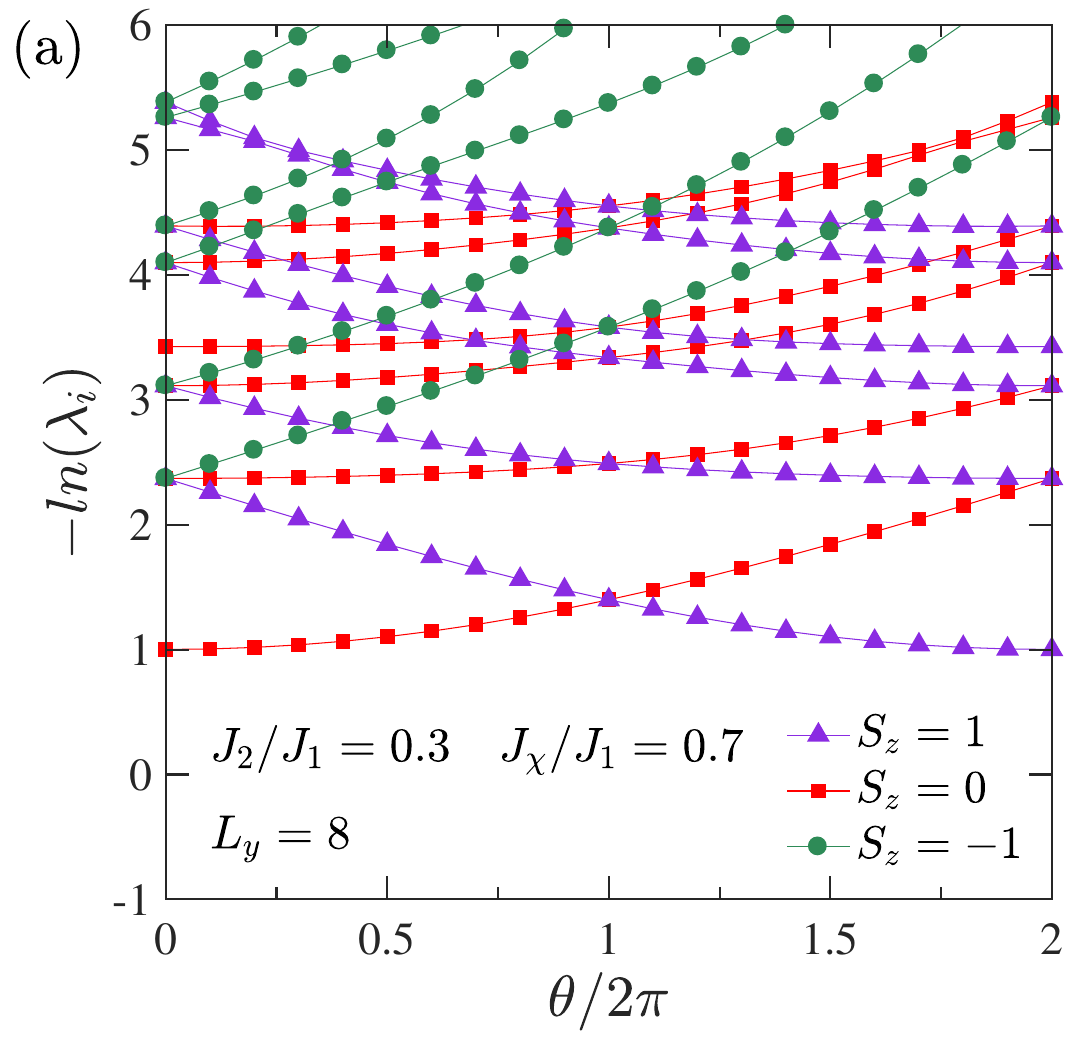}
\includegraphics[width = 0.85\linewidth]{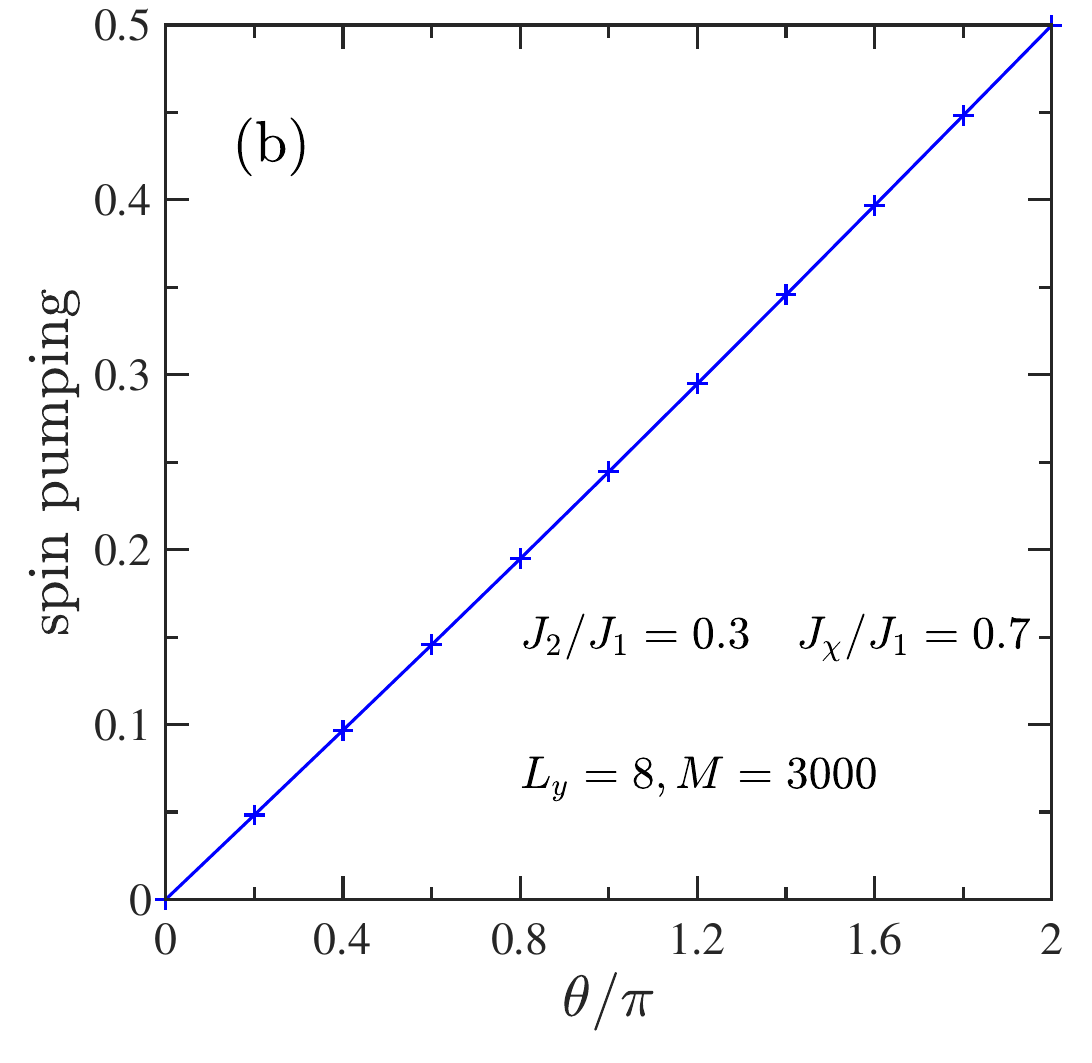}
\caption{Adiabatic flux insertion simulation and quantized spin Chern number for $J_2/J_1 = 0.3$, $J_\chi / J_1 = 0.7$ on the $L_y = 8$ cylinder.
(a) Entanglement spectrum flow with adiabatically inserted flux $\theta$, which is obtained by iDMRG simulation with $1000$ bond dimensions. The eigenvalues in the $S_z = 0, \pm 1$ sectors are labeled by different symbols. 
(b) Spin pumping with adiabatically inserted flux $\theta$, which is obtained by iDMRG simulation with $3000$ bond dimensions. In a period of inserted flux $\theta = 0 \rightarrow 2\pi$, a quantized magnetization moments $0.5$ is accumulated at one open boundary of the cylinder, indicating a fractionally quantized spin Chern number $C = 1/2$.}
\label{appfig:ES_flow_J2_03_Jx_07}
\end{figure}

In Fig.~\ref{fig:gap_TEE_entanglement} and Fig.~\ref{fig:ES_flow} of the main text, we have shown the DMRG results to identify the CSL state at $J_{\chi}/J_1 = 1.0$, $J_2 = 0$.
Here, we demonstrate additional DMRG data to support the CSL phase at $J_2 \neq 0$ in Fig.~\ref{appfig:ES_flow_J2_03_Jx_07}, for $J_2/J_1 = 0.3$, $J_\chi / J_1 = 0.7$ on the $L_y = 8$ cylinder.
The entanglement spectrum versus flux and the quantized spin Chern number $C = 1/2$ identify the CSL state with double-degenerate ground states.

\begin{figure}[ht]
\includegraphics[width = 0.85\linewidth]{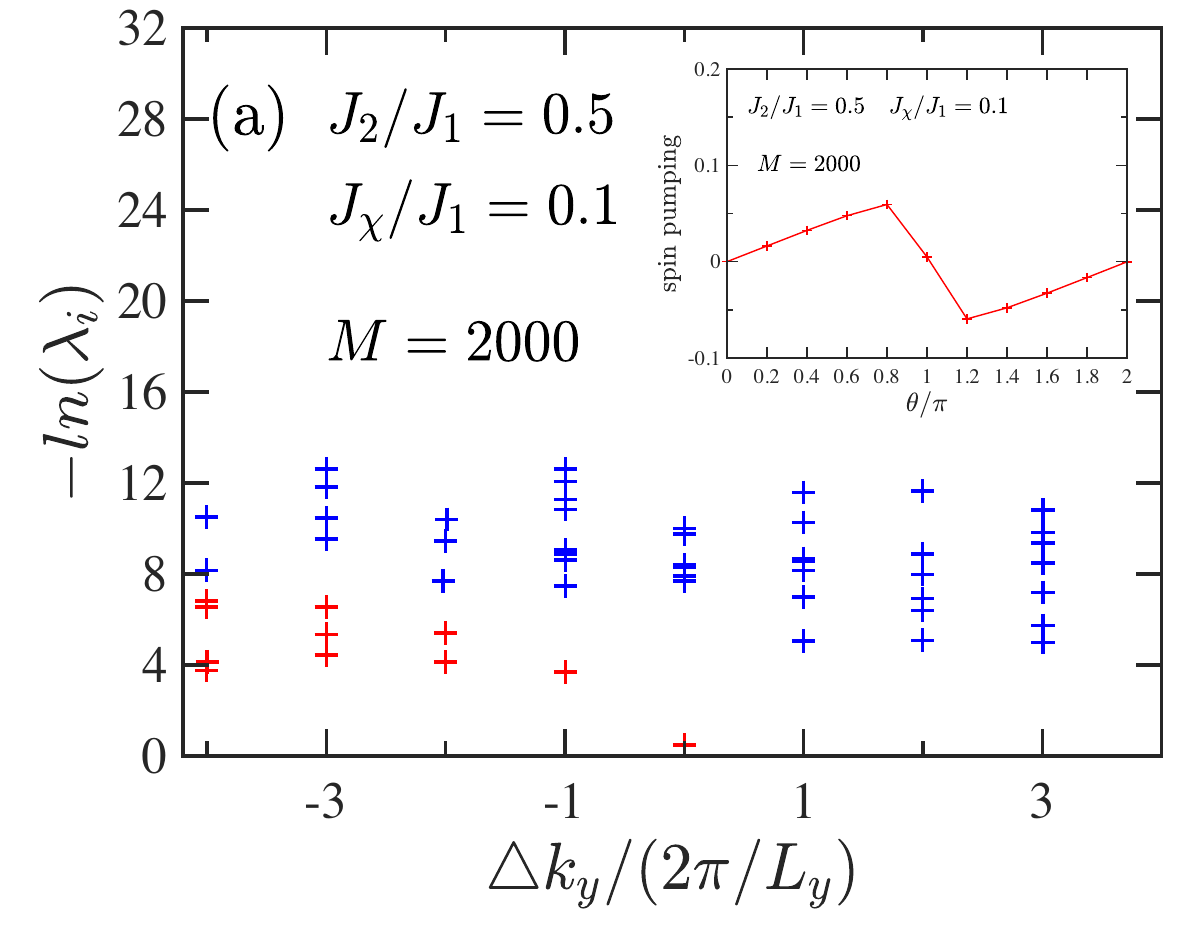}
\includegraphics[width = 0.85\linewidth]{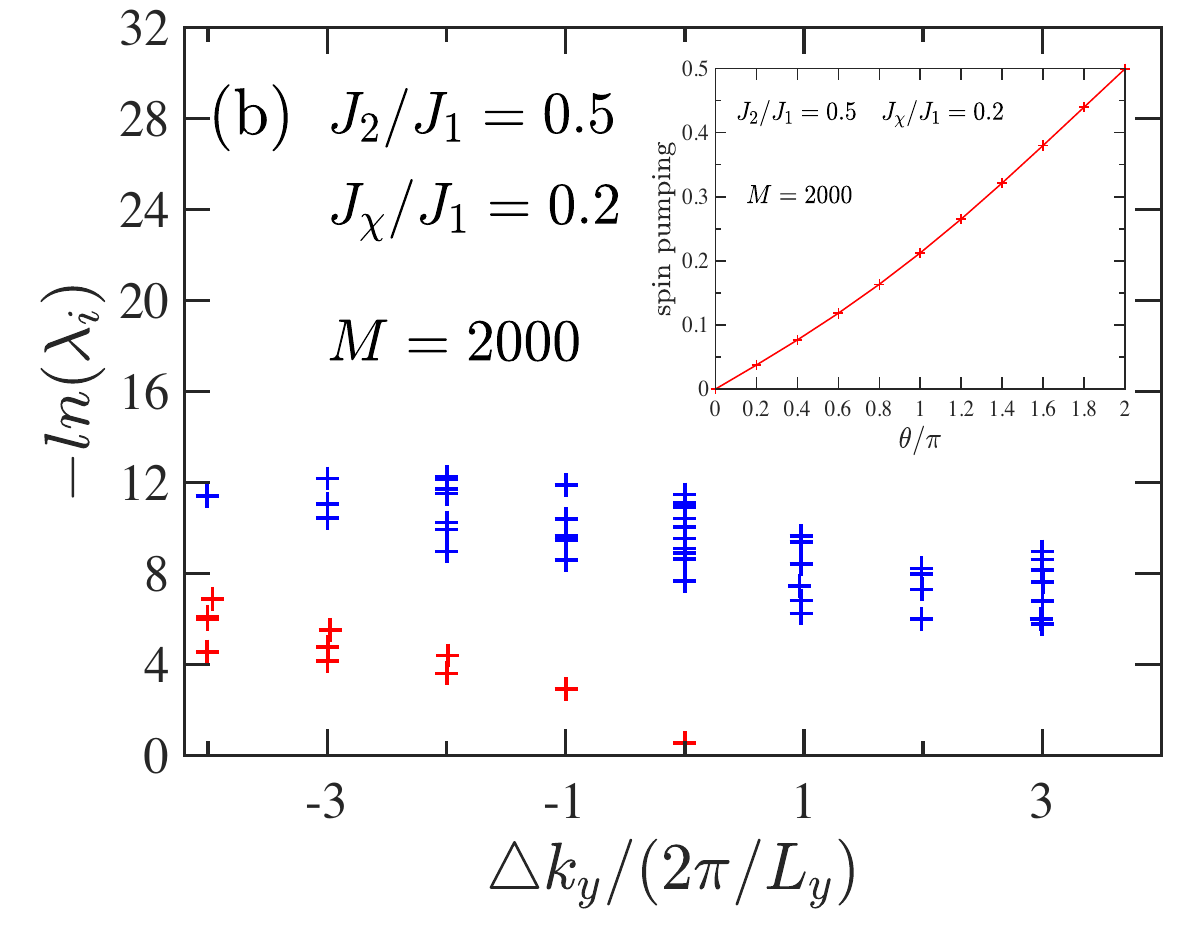}
\caption{Entanglement spectrum and adiabatic flux insertion simulations in the intermediate $J_2/J_1$ regime. (a) and (b) show the entanglement spectrum labeled by the quantum numbers of total spin $S^z = 0$ and relative momentum along the $y$ direction $\triangle{k_y}$ for the ground states of $J_2 / J_1 = 0.5$, $J_{\chi} / J_1 = 0.1$ and $0.2$, respectively, on the $L_y = 8$ cylinder systems. $\lambda_i$ are the eigenvalues of reduced density matrix. The results are obtained by keeping $2000$ U(1) states. The insets show the spin pumping with adiabatically inserted flux $\theta$.
The results at $J_\chi / J_1 = 0.2$ characterize the ground state as the CSL state.}
\label{appfig:ES_K_flow_J2_05_Jx}
\end{figure}

In the intermediate $J_2/J_1 \simeq 0.5$ regime, the $J_1$-$J_2$ model is a nonmagnetic state~\cite{LIU20221034}.
With growing chiral coupling, we find that the system will be easily driven to the CSL phase by a small chiral coupling.
In Fig.~\ref{appfig:ES_K_flow_J2_05_Jx}, we show the entanglement spectrum and adiabatic flux insertion results for $J_2 / J_1 = 0.5$, $J_\chi / J_1 = 0.1$ and $0.2$.
While the DMRG results at $J_\chi / J_1 = 0.1$ do not support a CSL, the CSL evidence is robust at $J_\chi / J_1 = 0.2$.
By using the similar simulations, we identify the CSL state in the intermediate $J_2/J_1$ regime with a small chiral coupling, as shown in Fig.~\ref{fig:model}.

\section{Analyses of magnetic order parameters}
\label{Appendix E: Analyses of magnetic order parameters}

\begin{figure}[h]
\includegraphics[width = 0.85\linewidth]{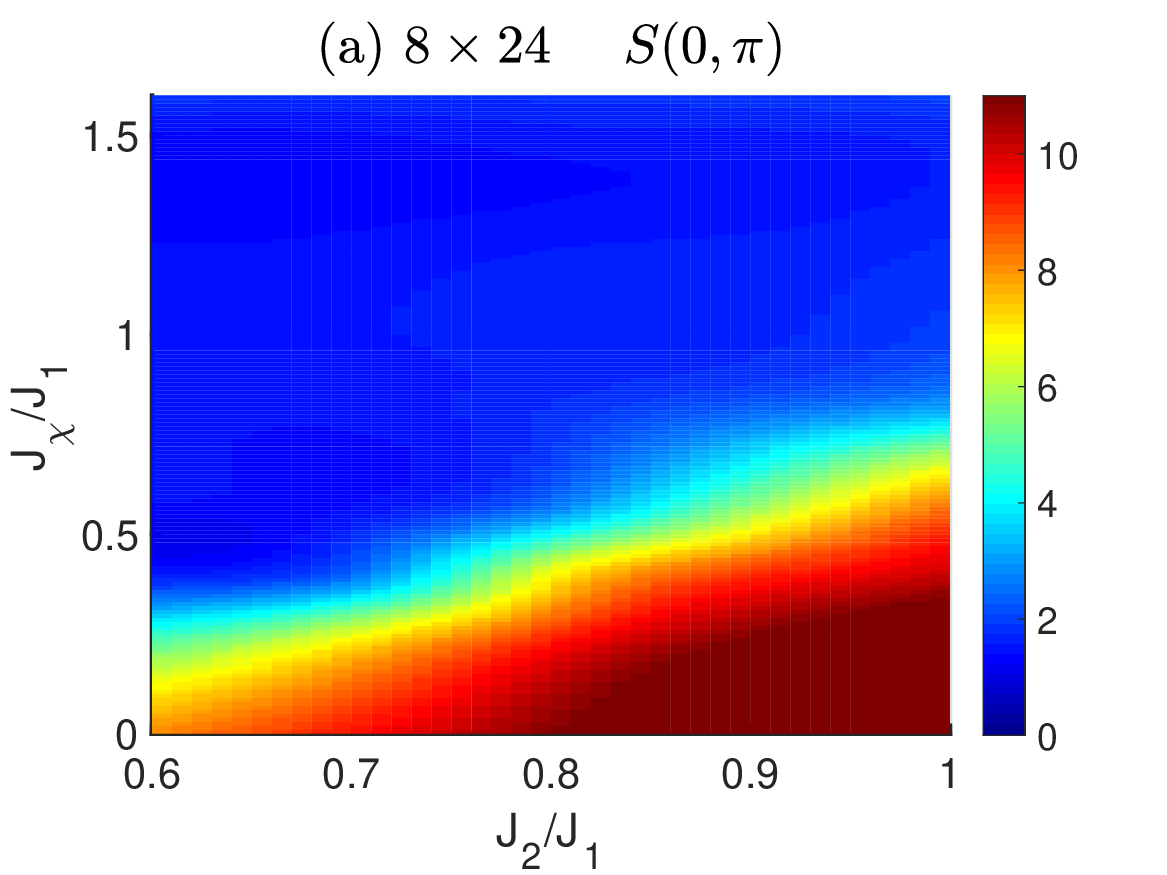}
\includegraphics[width = 0.85\linewidth]{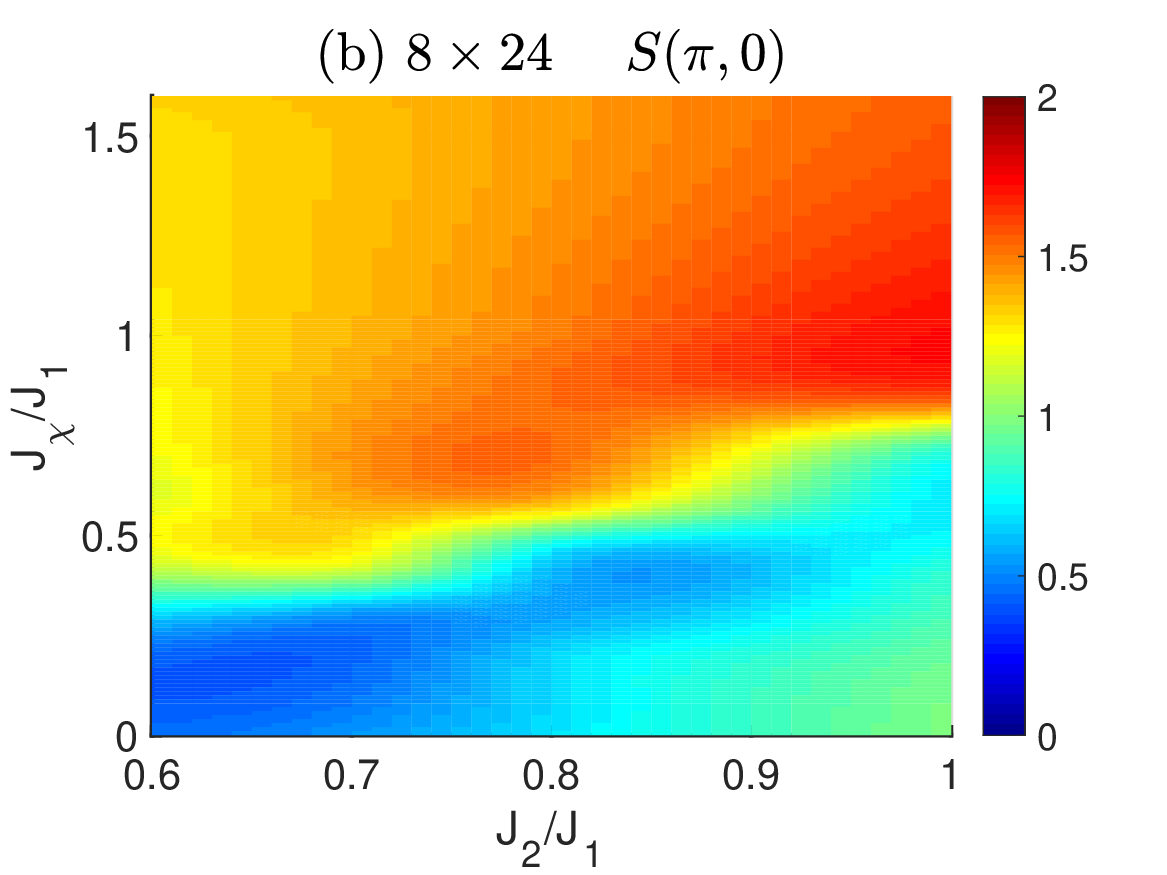}
\includegraphics[width = 0.85\linewidth]{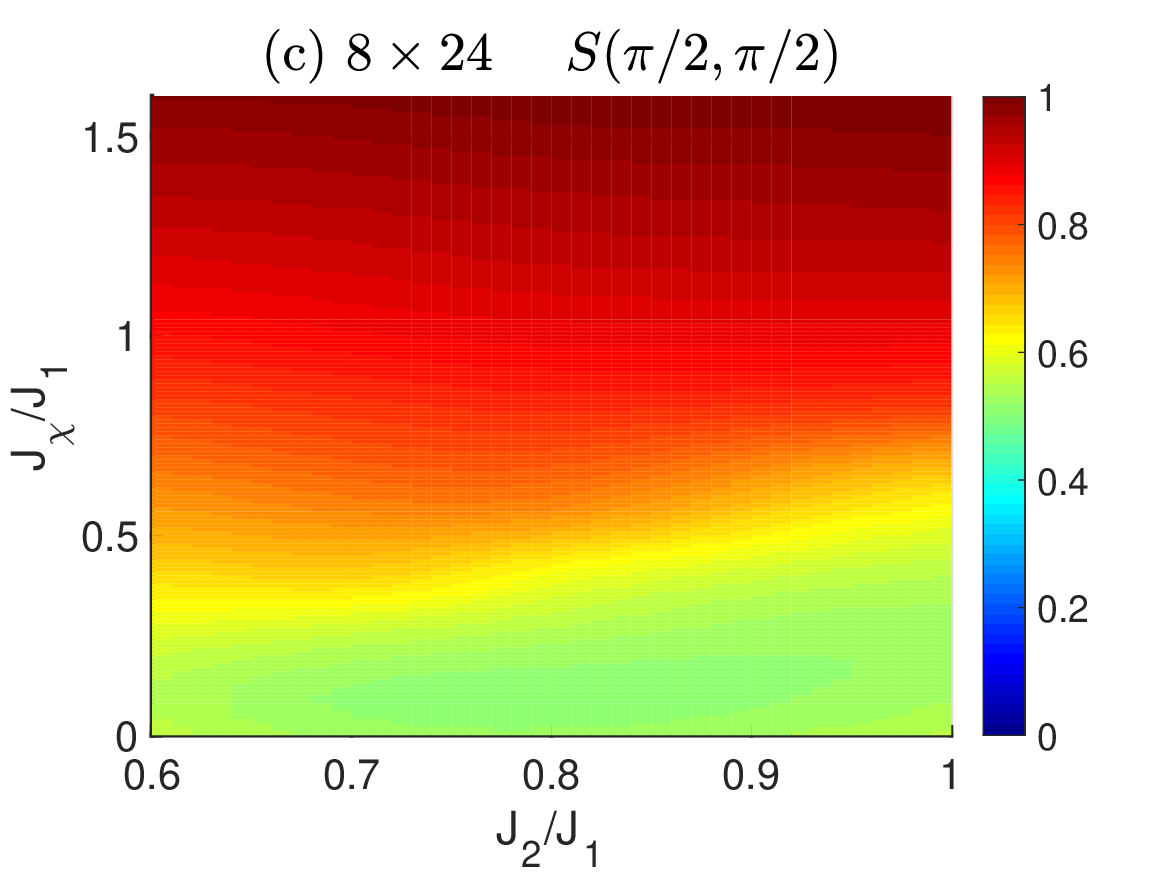}
\caption{Coupling dependence of spin structure factors on the eight-leg systems. (a)-(c) show the results $S(0,\pi)$, $S(\pi,0)$, and $S(\pi/2,\pi/2)$, respectively.}
\label{fig:CSS_Sq}
\end{figure}

In order to explore a possible weak CSS magnetic order in the magnetic disorder regime shown in Fig.~\ref{fig:model}, we have studied spin structure factor $S(0,\pi)$, $S(\pi,0)$ and $S(\pi/2,\pi/2)$ for $0.6 \leq J_2/J_1 \leq 1.0$ and $0\leq J_\chi/J_1 \leq 1.5$ on the $L_y=8$ cylinder, as shown in Fig.~\ref{fig:CSS_Sq}. 
In Fig.~\ref{fig:CSS_Sq}(a), $S(0,\pi)$ generally becomes weaker with increased chiral coupling. 
While $S(0,\pi)$ is strongly suppressed with vanishing stripe magnetic order, $S(\pi,0)$ shows the relatively stronger intensity for $0.8 \lesssim J_2/J_1 \lesssim 1.0$ and $0.7 \lesssim J_\chi/J_1 \lesssim 1.3$ as shown in Fig.~\ref{fig:CSS_Sq}(b).
Notice that in this parameter region, $S(0,\pi)$ and $S(\pi,0)$ have the similar magnitudes.
On the contrary, as demonstrated in Fig.~\ref{fig:CSS_Sq}(c), $S(\pi/2,\pi/2)$ is slightly enhanced with increased chiral coupling, in particular when the stripe order is suppressed.

\begin{figure}[h]
\includegraphics[width = 0.85\linewidth]{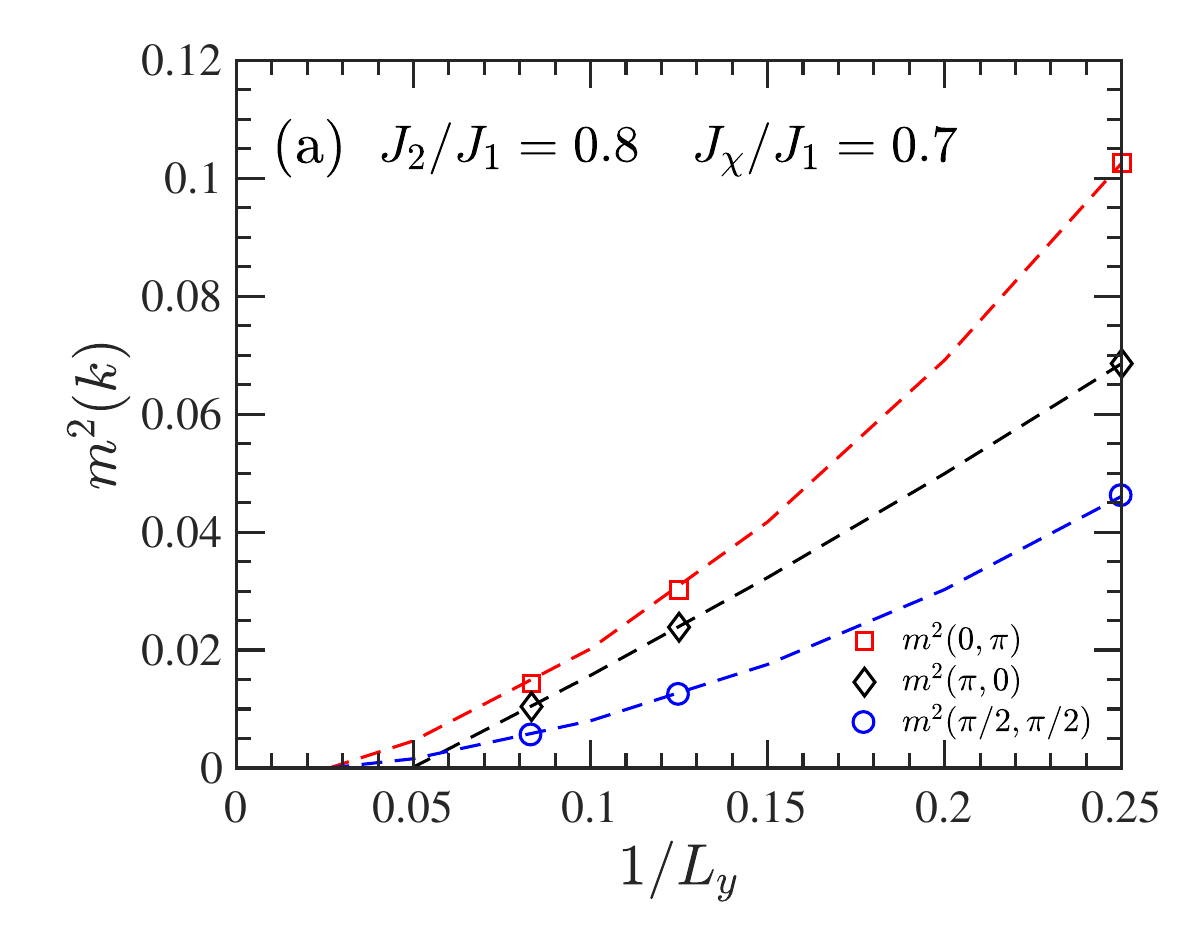}
\includegraphics[width = 0.85\linewidth]{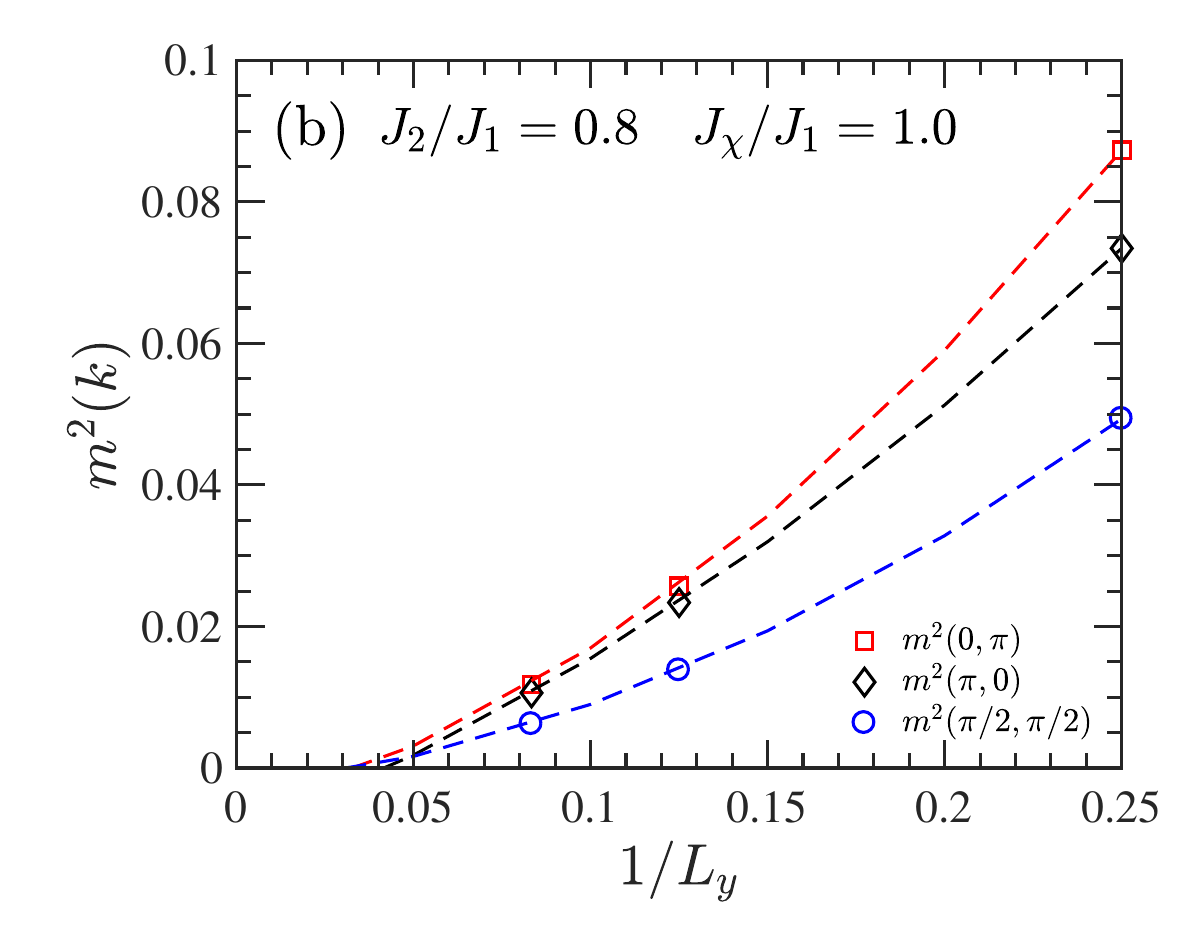}
\caption{Finite-size scaling of the magnetic order parameters $m^2(0, \pi)$, $m^2(\pi, 0)$ and $m^2(\pi/2, \pi/2)$ on the $L_y =4, 8, 12$ systems for (a) $J_2/J_1=0.8$, $J_\chi/J_1=0.7$, and (b) $J_2/J_1=0.8$, $J_\chi/J_1=1.0$. The results for $L_y=12$ are obtained by keeping $4000$ SU(2) multiplets.}
\label{fig:CSS_J2_08}
\end{figure}

To make finite-size scaling for magnetic order parameters, we choose three parameter points in the region where the spin structure factors $S(0,\pi)$, $S(\pi,0)$ and $S(\pi/2,\pi/2)$ are all relatively strong on the eight-leg system.
In Fig.~\ref{fig: CSS_m2}, we have shown the finite-size scaling of magnetic order parameters at $J_2/J_1 = J_\chi / J_1 = 1.0$.
Here, we show the results for $J_2/J_1=0.8$, $J_\chi / J_1 = 0.7, 1.0$ in Fig.~\ref{fig:CSS_J2_08}. 
Similar to the results in Fig.~\ref{fig: CSS_m2}, the finite-size scaling of the magnetic order parameters in Fig.~\ref{fig:CSS_J2_08} all decays fast with system width and is smoothly extrapolated to vanished, indicating the absent magnetic order.

\section{Coupling dependence of the dimer order decay length}
\label{Appendix F: VBC phase}

In Fig.~\ref{fig:bond_dimer}, we show the log-linear plot of the horizontal dimer order parameter $D_x$ for the pure $J_\chi$ model on different system widths, which demonstrates a strong tendency of the system to develop a bond dimer order.

\begin{figure}[ht]
\includegraphics[width = 0.85\linewidth]{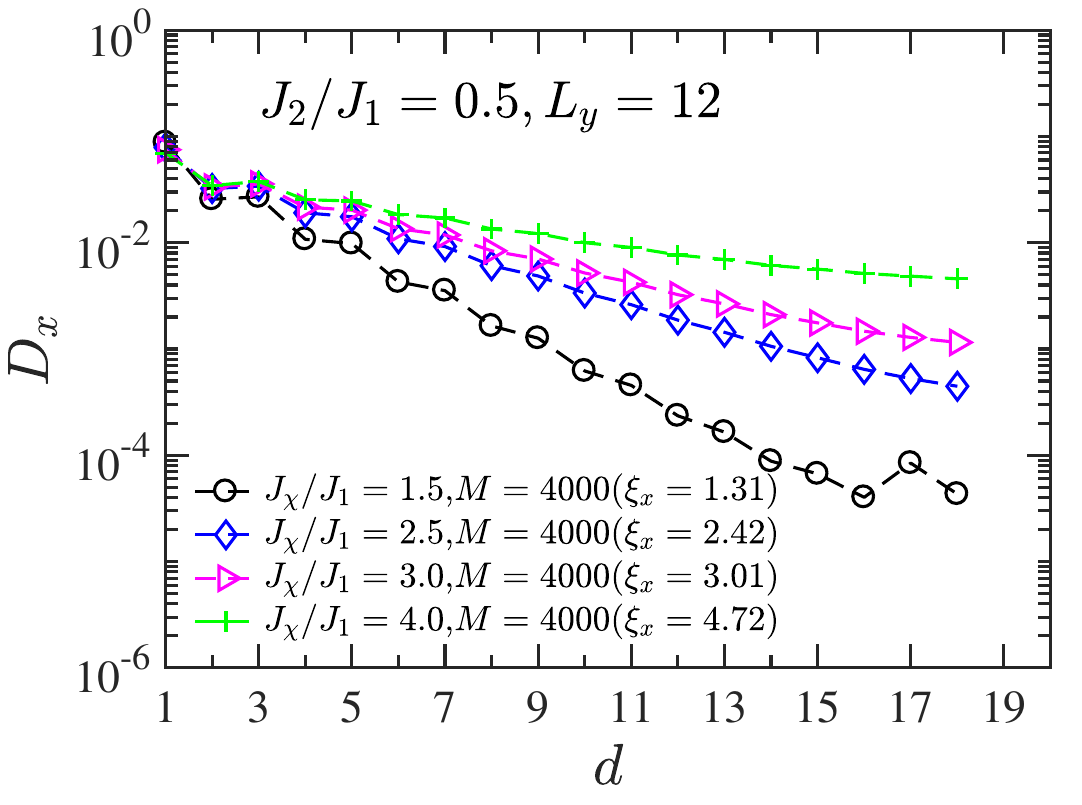}
\caption{Log-linear plot of the horizontal dimer order parameter $D_x$ versus the distance $d$ for $J_2/J_1 = 0.5$ and different $J_\chi/J_1$ on the $L_y = 12$ cylinder. $d$ is the distance of the bond from the open edge. The exponential fitting of $D_x \sim e^{-d/\xi_{x}}$ gives the decay length $\xi_x$. The results are obtained by keeping $4000$ SU(2) multiplets.
}
\label{fig:bond_dimer_J2_05}
\end{figure}

Here, we supplement the results at $J_2/J_1 = 0.5$ with different $J_\chi$ couplings on the $L_y = 12$ cylinder.
As shown in Fig.~\ref{fig:bond_dimer_J2_05}, the decay length $\xi_x$ of the dimer order parameter $D_x$ keeps growing with increased chiral coupling, which consistently suggests the stronger tendency to develop a dimer order at larger chiral coupling.

\clearpage
\bibliography{Square}
\end{document}